\documentclass[12pt]{article}
\usepackage{amsmath,epsfig,graphicx,verbatim}
\usepackage{epstopdf}
\usepackage{color}
\usepackage{subfigure}

\usepackage{here}
 \usepackage{graphicx}
  \usepackage{epsfig}

\usepackage{amsmath,bm}

\usepackage{cite}
\usepackage{lineno}

\include{epsf} 
\def\kt{\ensuremath{k_t}}
\newcommand{\as}{\alpha_\mathrm{s}}

\begin{document} 
\begin{titlepage}
\renewcommand{\thefootnote}{\fnsymbol{footnote}}
\begin{flushright}
  DESY 17 - 118 
\end{flushright}
\par \vspace{10mm}

\begin{center}
{\Large \bf Collinear and TMD 
Quark and Gluon Densities from 
\\[0.2cm]
  Parton Branching Solution 
of QCD Evolution Equations}
\end{center}

\par \vspace{2mm}
\begin{center}
{\bf F. Hautmann${}^{a, b, c},$ 
H. Jung${}^{d},$ 
A. Lelek${}^{d},$ 
V. Radescu${}^{e}$}  
and {\bf R. \v{Z}leb\v{c}\'{i}k$^{d}$\\
}

\vspace{5mm}

$^{a}$ Rutherford Appleton 
Laboratory, Chilton OX11 0QX 

$^{b}$ 
 Theoretical Physics 
Department, University of Oxford,  Oxford OX1 3NP

$^{c}$ Elementaire Deeltjes Fysica, Universiteit Antwerpen, B 2020 Antwerpen

$^{d}$ Deutsches 
Elektronen Synchrotron,  D 22603 Hamburg

$^{e}$  CERN,  CH-1211 Geneva 23

\vspace{5mm}

\end{center}

\par \vspace{2mm}
\begin{center} {\large \bf Abstract} \end{center}
\begin{quote}
\pretolerance 10000

We study parton-branching solutions of QCD evolution equations and present a method to construct both collinear and transverse momentum dependent (TMD) parton densities from this approach. We work with next-to-leading-order (NLO) accuracy in the strong coupling. Using the unitarity picture in terms of  resolvable and non-resolvable branchings,  we analyze the role of the soft-gluon resolution scale in the evolution equations. For longitudinal momentum distributions, we find agreement of our numerical calculations with existing evolution programs  
at the level of  better than 1\% over a 
range of five orders of magnitude both in evolution scale and in 
longitudinal momentum fraction. We make predictions for the evolution of 
transverse momentum distributions.  
We perform fits to the high-precision 
deep inelastic scattering (DIS) structure function measurements, and we present a set of 
NLO TMD distributions  based on the parton branching approach.

\end{quote}

\end{titlepage}

\setcounter{footnote}{1}
\renewcommand{\thefootnote}{\fnsymbol{footnote}}
\section{Introduction}
\label{sec:1}

Realistic comparisons of experimental data for hard processes 
 at high-energy 
hadron colliders with theoretical 
predictions based on 
QCD factorization 
formulas~\cite{Altarelli:1981ax,eswbook,Dissertori:2003pj,jcc-book} 
require Monte Carlo simulation 
via  parton shower event generators. While great progress 
has been achieved in the last 
decade on matching and merging 
methods~\cite{Sjostrand:2016bif,Nason:2012pr,Hoeche:2011zz} 
to combine parton showers with 
perturbative calculations 
through next-to-leading order (NLO),  
important open  questions still remain, 
both conceptual and technical, 
on the appropriate use of parton 
distribution functions in parton 
showers (see e.g.~\cite{Collins:2002ey,Collins:2005uv,Nagy:2014oqa,Nagy:2017ggp}) 
and on the treatment of the 
shower's transverse momentum kinematics  (see e.g.~\cite{Collins:2000qd,Hautmann:2012dw,Dooling:2012uw}). 
The relevance of these  effects is 
 known to increase with 
energy~\cite{mw92,hj-ang,Dooling:2014kia}, and they 
 thus 
constitute an important theme 
for physics at the Large Hadron 
Collider (LHC)  and at 
colliders of the next generation~\cite{Mangano:2016jyj}. 

Candidate approaches to tackle  
such questions in complex, multi-scale collider processes generally  
include  theoretical constructs  designed to extend  the concept   of collinear   parton density and decay functions, as in the case of 
Soft Collinear 
Effective  
Theory (SCET)~\cite{Stewart:2010qs,Alioli:2015toa,Bauer:2008qj} 
or of transverse momentum 
dependent (TMD) 
formalisms~\cite{Angeles-Martinez:2015sea,Hautmann:2012sh,Hautmann:2009zzb}.  
For instance, in  Ref.~\cite{hj-updfs-1312}   the TMD gluon density 
is determined,  based on the 
high-energy factorization~\cite{hef} and CCFM evolution 
equation~\cite{Catani:1989sg},   
from fits to the  high-precision deep inelastic scattering (DIS)  
data~\cite{Aaron:2009aa,comb-charm}, 
 and 
  used 
in a parton shower calculation~\cite{Dooling:2014kia}
to make predictions for $W$-boson + jets hadro-production, which can be   
compared   with 
LHC experimental measurements~\cite{atlas-w-jets,cms-w-jets}. 

Although the analysis  
in~\cite{Dooling:2014kia,hj-updfs-1312} proves to be successful 
in achieving a meaningful  TMD description of both 
DIS  and Drell-Yan  measurements, 
it is based on a TMD form of factorization valid 
at high energy, and  requires a 
matching method 
(provided by the CCFM equation)
to include non-asymptotic contributions at collider energies.  Because 
of the high-energy expansion, 
the method is predominantly 
sensitive to the gluon density, 
and quark contributions enter 
systematically at subleading 
orders. 

In Ref.~\cite{Hautmann:2017xtx}  we have proposed  
 a   different approach,  
based on solving coupled 
quark and gluon 
 DGLAP~\cite{dglaprefs}  evolution  equations
using parton branching  
methods, and determining from 
this  
both collinear  (integrated over transverse momenta, iTMD) 
and TMD parton 
densities. Rather than starting 
from high-energy resummed 
 equations, this work    relies  
on  renormalization group evolution 
equations.  
The approach uses  the unitarity 
formulation of  these  evolution 
equations which forms the basis of 
parton showering Monte Carlo 
simulation~\cite{eswbook,Webber:1986mc}. 
In this sense, it is close  in spirit 
to the works in 
Refs.~\cite{Jadach:2003bu,GolecBiernat:2006xw,GolecBiernat:2007pu,Jadach:2005bf,Stoklosa:2007zz,jada09,Jadach:2008nu,Jadach:2012vs,Jadach:2015mza},  in     Refs.~\cite{Tanaka:2001ih,Kurihara:2002ne,Tanaka:2003ck,Tanaka:2003gk,Kato:1991fs,Tanaka:2005rm,Tanaka:2014yxa}, and in Refs.~\cite{Hoche:2017iem,Hoche:2017hno}.    
Ref.~\cite{Hautmann:2017xtx} shows 
that the evolution of parton distribution 
functions can be calculated, including 
the transverse momentum dependence, 
from a parton branching approach,  provided infrared contributions  
to evolution 
are treated by a method which takes into account 
consistently soft gluon emissions,  near the endpoint for lightcone 
momentum fractions $ z \to 1$,  not just at inclusive level but at exclusive level. In particular, 
Ref.~\cite{Hautmann:2017xtx} 
shows that this can be done 
by using a  finite  
soft-gluon resolution scale  in the evolution equations. 

In this paper we provide 
details of the approach set out 
in~\cite{Hautmann:2017xtx}, we show that it can be applied to higher  accuracy order-by-order in the strong coupling $\alpha_s$, and we present   numerical results at the 
next-to-leading order (NLO).  
 Further we present the results of 
fits, based on this approach, to the high-precision DIS 
data~\cite{Abramowicz:2015mha}.  
First results from this work have appeared  in~\cite{Lelek:2016hro}.

The paper is organized as follows. In Sec.~\ref{sec:2} we 
describe the main elements of the 
parton-branching 
formulation of the coupled 
QCD evolution equations.  In 
Sec.~\ref{sec:3} we 
present the numerical Monte Carlo solution 
of the coupled quark and gluon evolution 
 equations at NLO. We compare   collinear parton 
density functions obtained by our parton-branching   
solution  with results 
obtained via the evolution package 
{\sc Qcdnum}~\cite{Botje:2010ay,qcdnum-pre1,qcdnum-pre2}. 
In Sec.~\ref{sec:4} we  illustrate  an  application of our method 
by performing a  fit to the high-precision DIS data~\cite{Abramowicz:2015mha}.  
For the fit we use an updated version of the 
program~\cite{Hautmann:2014uua}  within 
the   \verb+xFitter+  open-source QCD fit platform~\cite{Alekhin:2014irh}.  
By the method of the present paper we are able to extend the 
fit~\cite{hj-updfs-1312} to precision DIS data significantly 
toward  higher $x$ and higher $Q^2$.
In Sec.~\ref{sec:5} we turn to TMD parton 
density functions~\cite{Angeles-Martinez:2015sea,tmdplott}. We discuss the 
identification of the transverse momentum in 
the initial-state 
parton distribution in terms of the 
  shower's  kinematics and evolution variable. 
We present 
a new set of  quark 
and gluon TMDs  
including  
NLO evolution kernels.   
  We give conclusions 
in Sec.~\ref{sec:6}.

\section{Unitarity approach 
to QCD evolution equations}
\label{sec:2}

In this section  we give  the main elements 
of the parton-branching approach to   the evolution equations. 
We introduce a soft-gluon resolution scale into the 
renormalization group evolution equations, and    describe   
 resolvable and non-resolvable 
emissions. We discuss  
 the relationship of our results 
with 
the angular-ordered, coherent 
branching~\cite{Dokshitzer:1987nm,bcm83,Marchesini:1987cf,Catani:1990rr}
and the behavior of the endpoint 
$z \to 1$ region 
in transverse momentum 
distributions~\cite{Hautmann:2000cq,Hautmann:2007uw}. 
We construct an iterative Monte 
Carlo solution of the evolution equations,  and apply it 
to the case of collinear and TMD 
parton densities.

\subsection{The renormalization 
group evolution} 
\label{subsec:2a} 

The  renormalization group  evolution of parton distribution functions 
can be written in terms of parton splitting processes as follows 
\begin{equation}
\frac{\partial \, f_{a}(x, \mu^{2})}{\partial \ln \mu^{2}} =  \sum_{b} \int_{x}^{1}
\frac{dz}{z} P_{ab}(\as (\mu^{2}), z) \;f_{b}(x/z,\mu^{2}) \;,
\label{dfa}
\end{equation}
where $f_{a}(x, \mu^{2})$ 
 are  parton distributions   for $a = 1 , \dots , 2 N_f + 1 $ species of partons   
(with $N_f$  the number of quark  flavors)   
 as functions of longitudinal 
momentum fraction $x$ and  
evolution mass scale $\mu$, and  
$P_{ab} (\as ,z)$ are  the DGLAP 
splitting functions,  
depending on the running coupling 
$\as$ and the splitting variable $z$, 
and 
computable as a  perturbation 
series expansion 
\begin{equation}
P_{ab} (\as ,z) = \sum^{\infty}_{n=1} \left( \frac{\as}{2\pi}
\right)^{n} P_{ab}^{(n-1)}(z)\;.
\label{Pab}
\end{equation}
We will work with the 
momentum-weighted parton 
distribution functions  
$\widetilde f_{a}$,  
\begin{equation}
{\widetilde f}_{a} (x,\mu^{2}) \equiv x f_{a} (x,\mu^{2}) \;\;,
\label{tildef}
\end{equation}
for which the evolution equations 
read 
\begin{equation}
\label{evapp}
\frac{\partial \;{\widetilde f}_a(x,\mu^2)}{\partial \ln \mu^2} = \sum_b \int_x^1 {dz} \;
P_{ab}(\as(\mu^2),z) \;{\widetilde f}_b({x/z},\mu^2) \;\;.  
\end{equation} 

In the physical picture of Eq.~(\ref{evapp}), a finite resolution scale 
in the transverse distance between emitted partons implies, by 
energy-momentum 
conservation,  that one cannot 
resolve 
partons radiated with 
longitudinal momentum fractions closer to $ z = 1$ than a 
certain cut-off value, 
$ z > z_M $ with $ 1-z_M \sim 
{\cal O} ( \Lambda_{\rm{QCD}} / 
\mu )$,   
where $\mu$ is of the order of 
the hard-scattering scale and $  \Lambda_{\rm{QCD}} \approx 
1$~fm$^{-1}$ is the natural scale 
of the strong interactions. 
Removing non-resolvable    radiative  contributions from the evolution, on 
the other hand,  leads to a violation 
 of unitarity. 
The key idea of the parton branching 
method is to restore unitarity by  recasting 
 the evolution equations in terms of 
no-branching probabilities (Sudakov 
form factors) and 
real-emission branching 
 probabilities. 
 We will introduce 
the resolution scale parameter 
$z_M$ formally into the evolution 
equations in Sec.~\ref{subsec:2c},    
and  describe the unitary branching method  in the subsequent  sections.

To set up our formalism,  
we decompose the splitting functions 
$P_{ab} (\as ,z)$ as 
\begin{equation}
P_{ab} (\as ,z) = 
D_{ab} (\as) \delta ( 1 - z ) 
+ K_{ab} (\as) \ 
{ 1 \over { ( 1 - z )_+ }} 
+ R_{ab} (\as ,z)
\;, 
\label{decompPab}
\end{equation}
where the plus-distribution 
${ 1 / { ( 1 - z)_+ }} $ 
is defined for any test function 
$\varphi$ as 
\begin{equation}
\int_0^1 { 1 \over { ( 1 - z)_+ }}  \ 
\varphi ( z )  \ dz  = 
\int_0^1 { 1 \over {  1 - z }}  \ [ 
\varphi ( z )  - \varphi ( 1 ) ] \  dz
\;. 
\label{plusdistrdef}
\end{equation}
Eq.~(\ref{decompPab}) 
provides a classification of the 
singular behavior of the 
 splitting functions 
$P_{ab} (\as ,z)$ in the non-resolvable radiation 
 region $ z \to 1$.  
It decomposes 
the splitting functions into the 
 $\delta (1-z)$ distribution, 
the ${ 1 / { ( 1 - z)_+ }} $ distribution, 
and the function  $R (\as ,z)$ which 
contains logarithmic terms in 
 $\ln (1-z)$ and analytic terms for 
$z \to 1$.  The 
$\delta (1-z)$ 
and ${ 1 / { ( 1 - z)_+ }} $
contributions to  
splitting functions    are 
diagonal in flavor, 
\begin{equation}
D_{ab} (\as) = \delta_{a b } 
d_a (\as)  
\, , \;\; 
K_{ab} (\as) = 
\delta_{a b } 
k_a (\as)  
\,  
\label{flavdiag}
\end{equation}
(no summation  
over repeated indices).  
The 
functions 
$ D_{ab} $ and  $ K_{ab} $, 
or equivalently 
$d_{a}$ 
and $k_{a}$,  and the functions 
$ R_{ab} $ 
in Eq.~(\ref{decompPab}) 
have the perturbation series expansions 
\begin{equation}
d_{a} (\as) = \sum^{\infty}_{n=1} \left( \frac{\as}{2\pi}
\right)^{n} d_{a}^{(n-1)}  
\, , \;\; 
k_{a} (\as) = \sum^{\infty}_{n=1} \left( \frac{\as}{2\pi}
\right)^{n} k_{a}^{(n-1)}  
\, , 
\label{Aab}
\end{equation}
\begin{equation}
R_{ab} (\as ,z) = \sum^{\infty}_{n=1} \left( \frac{\as}{2\pi}
\right)^{n} R_{ab}^{(n-1)}(z)\;.
\label{Rab}
\end{equation}

The treatment which we 
develop in this section 
only relies on the decomposition 
in Eq.~(\ref{decompPab}) and
is valid  at any order in $\as$. 
In practical applications 
one takes a given truncation 
of the expansions in 
Eqs.~(\ref{Aab}), (\ref{Rab}). 
The numerical results in 
Secs.~\ref{sec:3}, \ref{sec:4} 
and \ref{sec:5} are based 
on the expansion to NLO (i.e., 
$n = 2$ in Eqs.~(\ref{Aab}), (\ref{Rab})).

Charge conjugation and $SU(N_f)$  flavor symmetries  
imply that the splitting functions  $P_{a b}$   obey the following relations 
to all orders,  
\begin{eqnarray}
\label{adpr}
P_{q_ig}=P_{{\bar q}_ig} \equiv P_{qg} \;\;\;\;&,& \;\;\;\;\;
P_{gq_i}=P_{g{\bar q}_i} \equiv P_{gq} 
\;\;,    
\nonumber\\
P_{q_iq_j}=P_{{\bar q}_i{\bar q}_j} \equiv P_{qq}^{NS} \delta_{ij} +
P_{qq}^{S} \;\;&,& \;\;\;
P_{q_i{\bar q}_j}=P_{{\bar q}_iq_j} \equiv P_{q{\bar q}}^{NS} 
\delta_{ij} + P_{q{\bar q}}^{S} \;\;, 
\end{eqnarray}
where the superscripts $NS$ and $S$ stand respectively for non-singlet and singlet. 
Therefore,  $P_{a b}$ has  
three independent 
quark-gluon or gluon-gluon 
components 
($P_{qg}$, $P_{gq}$ and $P_{gg}$) 
and 
four independent  
quark-quark 
components 
(the $NS$ components 
$P_{qq}^{NS}$, $
P_{q{\bar q}}^{NS}$ and the $S$ components 
$ P_{qq}^{S}  $, $P_{q{\bar q}}^{S}$).\footnote{The 
independent quark-quark 
 components can alternatively be 
 taken~\cite{Altarelli:1981ax,Catani:1994sq,cfpref1}   
  to 
 be the 
 three 
 which correspond to the 
 three linear combinations  diagonalizing  
the evolution of non-singlet distributions, plus the 
one  which controls the 
evolution of  the singlet  quark   distribution coupled 
to gluons.}  
 
 

In the next section we give explicit expressions at one-loop and two-loop orders for the 
$D_{ab}$, $K_{ab}$ and  $R_{ab}$ terms  in 
Eq.~(\ref{decompPab}). 

\subsection{Expansion in 
powers of 
{\boldmath$\as$}}
\label{subsec:2b}

At one-loop order    the 
coefficients of the perturbative expansions (\ref{Aab}),(\ref{Rab}) 
for    $d_{a}$,  
 $k_{a}$ and  $ R_{ab} $  can be read  from~\cite{dglaprefs}. 
At this order, one has $
P_{q{\bar q}}^{NS} = P_{qq}^{S}  = P_{q{\bar q}}^{S}  = 0$, so that 
all quark-quark components are degenerate. 
The one-loop expressions for  $d_{a}$,  
 $k_{a}$ and  $ R_{ab} $ are given by 
\begin{equation}
 d^{(0)}_q =  {3 \over 2} \, C_F \;\;  
\; , \;\; d^{(0)}_g = 
{11 \over 6} \, C_A - 
{2 \over 3} \, T_R \, N_f
\; , 
\label{oneloopA}
\end{equation}

\begin{equation}
k^{(0)}_q = 2 \ C_F 
\; , \;\; k^{(0)}_g = 
2 \ C_A   
\;  
\label{oneloopK}
\end{equation}
and 
\begin{eqnarray}
\label{oneloopR}
R_{g g}^{(0)} (z ) &=& 2 \, C_A \, \left[  
 {{1-z} \over z} + z (1 - z)  -1 \right]  \;\;,
\nonumber\\
R_{g q_i}^{(0)} (z ) &=& 
R_{g {\bar q}_i}^{(0)} (z ) = 
 C_F \, 
 {{1+ (1-z)^2} \over z} \;\;, 
\nonumber\\
R_{ q_i g}^{(0)} (z ) &=& 
R_{ {\bar q}_i g}^{(0)} (z ) = T_R \, \left[ z^2 + (1-z)^2 \right] \;\;, \\
R_{ q_i q_j}^{(0)} (z ) &=& 
R_{ {\bar q}_i {\bar q}_j}^{(0)} (z ) = 
- C_F \, (1+z)  \, \delta_{i j} \;\;, 
\;\;\;\;
R_{ { q}_i {\bar q}_j}^{(0)} (z ) = 
R_{ {\bar q}_i { q}_j}^{(0)} (z ) = 0 \;\;, \nonumber
\end{eqnarray}
where the $SU(N_c)$ color factors   (with $N_c =3 $  
the  number of colors) 
are given by 
\begin{equation}
\label{colorfactor}
C_A = N_c \;\;\;, \;\;\;\;\; C_F = {{N_c^2 - 1} \over { 2 \, N_c}} \;\;\;, 
\;\;\;\;\; 
{\mbox {\rm Tr}} \,(t^k t^m) = 
\delta^{ k m } \, T_R  = {1 \over 2} \, 
\delta^{ k m}  \;\;\;.
\end{equation}

At two-loop order  
the perturbative coefficients for   $d_{a}$,  
 $k_{a}$ and  $ R_{ab} $ start to depend on the renormalization  scheme.  
In the ${\overline {\rm{MS}} }$ scheme  the results 
can be read 
from~\cite{cfpref1,cfpref2}. 
At  
 the level of two loops  
also 
$
P_{q{\bar q}}^{NS}$, $ P_{qq}^{S} $, $ P_{q{\bar q}}^{S}  $  are nonvanishing so that 
the degeneracy of the quark splitting functions 
 is   lifted. However  
 a residual degeneracy remains  because 
 $ P_{qq}^{S} = P_{q{\bar q}}^{S}  $ at this 
order.\footnote{The
degeneracy is fully lifted starting 
at three-loop order~\cite{mochetal-NS,mochetal-S}.}  
The two-loop coefficients for $d_{a}$ and 
 $k_{a}$ 
are given by 
\begin{eqnarray}
 d^{(1)}_q &=&  
C_F^2 \left( {3 \over 8} -  
{\pi^2 \over 2} + 6 \ \zeta (3) \right) 
+ C_F C_A \left( 
{17 \over 24} +   
{{11 \pi^2} \over 18} - 3 \ \zeta (3)
 \right) 
-  C_F T_R N_f \left( 
{1 \over 6} +   
{{2 \pi^2} \over 9} 
 \right)   \; , 
\nonumber\\ 
 d^{(1)}_g &=&  
C_A^2 \left( {8 \over 3} + 3 \ 
 \zeta (3) \right) 
- {4 \over 3} \ C_A T_R N_f 
-  C_F T_R N_f   \; , 
\label{twoloopA}
\end{eqnarray}
where $\zeta$ is the Riemann zeta function, and 
\begin{eqnarray}
 k^{(1)}_q &=&   
2 \ C_F \ \Gamma 
   \; , 
\;\; k^{(1)}_g =   
2 \ C_A \ \Gamma 
   \; , 
\nonumber\\ 
{\rm{where}} \;\;   \Gamma  &=&  
 C_A \left( 
{ 67 \over 18 } - {\pi^2 \over 6} 
\right) - T_R N_f {10 \over 9 } 
 \; . 
\label{twoloopK}
\end{eqnarray}
The expressions for the two-loop coefficients for   
the functions $R_{ab}$ are lengthier,  
and are given in  Appendix A. 

We will use these  
expansions through two loops 
  for the numerical calculations in 
Secs.~\ref{sec:3}, \ref{sec:4},  
 \ref{sec:5}. 

\subsection{Resolvable and 
non-resolvable emissions} 
\label{subsec:2c} 

We  now 
introduce the soft-gluon 
resolution 
 parameter $z_M$ 
 into the evolution equations (\ref{evapp}), by 
  splitting  the 
integration range on the 
right hand side into the 
resolvable 
($ z < z_M$)  and non-resolvable 
($ z > z_M$) regions, where 
 $ 1-z_M \sim 
{\cal O} ( \Lambda_{\rm{QCD}} / 
\mu )$. 
In each region, we 
use the decomposition~(\ref{decompPab}) in 
the evolution 
equations. We   
  include terms through 
$(1-z_M)^0$ but neglect power-suppressed contributions $ {\cal O} 
(1 - z_M) ^n $, $ n \geq 1$.

Consider first the endpoint 
$ z \to 1 $ contribution 
from the 
$K_{ab}$ term in 
Eq.~(\ref{decompPab}). 
Using 
Eq.~(\ref{plusdistrdef}),   
we rewrite this  
as 
\begin{eqnarray}
\label{rewr1}
&& \sum_b \int_x^1 {dz} \;
K_{ab}(\as(\mu^2))  \ 
{ 1 \over { ( 1 - z )_+ }} 
 \;  {\widetilde f}_b({x/z},\mu^2) 
\\ 
&& 
= 
 \sum_b \int_x^1 {dz} \;
K_{ab}(\as(\mu^2))  \ 
{ 1 \over {  1 - z  }} 
 \;  {\widetilde f}_b({x/z},\mu^2) 
- \sum_b \int_0^1 {dz} \;
K_{ab}(\as(\mu^2))  \ 
{ 1 \over {  1 - z  }} 
 \;  {\widetilde f}_b({x},\mu^2) 
\;\;   .  
\nonumber 
\end{eqnarray}
In the region 
$ 1> z > z_M$ 
we  expand the   
momentum-weighted parton 
density   as 
\begin{equation} 
\label{deriv-expan-f} 
{\widetilde f}_b({x/z},\mu^2) = 
{\widetilde f}_b({x},\mu^2)
+  (1-z) {{\partial {\widetilde f}_b } \over {\partial \ln x }} ({x},\mu^2) + {\cal O}(1-z)^2 \;\; . 
\end{equation}
 Then we see that the contribution 
to Eq.~(\ref{rewr1}) from the 
non-resolvable region is of order 
${\cal O}(1-z_M)$, and thus, 
up to ${\cal O}(1-z_M)
$, 
we have   
\begin{eqnarray}
\label{rewr3}
&& \sum_b \int_x^1 {dz} \;
K_{ab}(\as(\mu^2))  \ 
{ 1 \over { ( 1 - z )_+ }} 
 \;  {\widetilde f}_b({x/z},\mu^2) 
\\ 
&=& 
 \sum_b \int_x^{z_M} {dz} \;
K_{ab}(\as(\mu^2))  \ 
{ 1 \over {  1 - z  }} 
 \;  {\widetilde f}_b({x/z},\mu^2) 
- \sum_b \int_0^{z_M} {dz} \;
K_{ab}(\as(\mu^2))  \ 
{ 1 \over {  1 - z  }} 
 \;  {\widetilde f}_b({x},\mu^2) 
 \;\; .   
\nonumber
\end{eqnarray}

Next, we consider the contributions 
 to the evolution equations 
(\ref{evapp})  from the other two 
terms, $D_{ab}$ and $R_{ab}$, 
in Eq.~(\ref{decompPab}). 
The $R_{ab}$ contribution can be combined with the 
first term on the right hand side 
of Eq.~(\ref{rewr3}) to yield 
a  
contribution to the evolution 
 proportional to 
${\widetilde f}_b({x/z},\mu^2)$.
 The 
 $D_{ab}$ contribution can be combined with the 
second term on the right hand side 
of Eq.~(\ref{rewr3}),  using the 
$\delta (1-z)$,  to yield  
a  contribution to the evolution  proportional to ${\widetilde f}_b({x},\mu^2)$. 
Further, we use that 
$R_{ab}$ has no 
power divergences $(1-z)^{-n}$ 
and is at most 
logarithmic for $ z \to 1$, so 
that 
the integration over 
$R_{ab}$ for $ z > z_M$ 
gives ${\cal O}(1-z_M)$. 
Thus, we can write 
\begin{eqnarray}
\label{rewr4zeroth}
&& \frac{\partial \;{\widetilde f}_a(x,\mu^2)}{\partial \ln \mu^2}  =  \sum_b 
\int_x^{z_M} {dz} \;
\left( K_{ab} (\as(\mu^2)) \ 
{ 1 \over {  1 - z  }} 
+ R_{ab} (\as(\mu^2) , z) 
\right) 
\;{\widetilde f}_b({x/z},\mu^2) 
\nonumber\\ 
 & + & \sum_b 
\left\{  \int_x^{1}  
D_{ab}(\as(\mu^2))  \ 
\delta (  1 - z ) \  dz 
-  \int_0^{z_M} 
K_{ab}(\as(\mu^2))  \ 
{ 1 \over {  1 - z  }}  \ dz 
 \right\} 
  {\widetilde f}_b({x},\mu^2) \;\; .  
\end{eqnarray} 
 
The first line in 
Eq.~(\ref{rewr4zeroth}) contains 
contributions to evolution from 
real parton emission,  
while the second line 
contains contributions from 
virtual  corrections. 
It is 
convenient to 
  define  the kernels  in the bracket 
of the first line 
as the real-emission branching probabilities $P_{ab}^{(R)} (\as ,z)$, 
\begin{equation}
P_{ab}^{(R)} (\as ,z) = 
 K_{ab} (\as) \ 
{ 1 \over {  1 - z  }} 
+ R_{ab} (\as ,z)
\; .  
\label{realPab}
\end{equation}
That is, the 
real-emission branching 
probabilities $P_{ab}^{(R)} (\as ,z)$ 
are obtained 
 from the splitting functions 
$P_{ab} (\as ,z)$ in  
Eq.~(\ref{decompPab})  
by subtracting the $\delta (1-z)$ 
terms   and 
replacing the plus-distribution 
${ 1 / { ( 1 - z)_+ }} $ 
by ${ 1 / { ( 1 - z) }} $. 
So we have 
\begin{eqnarray}
\label{rewr4first}
&& \frac{\partial \;{\widetilde f}_a(x,\mu^2)}{\partial  \ln \mu^2}  =  \sum_b 
\int_x^{z_M} {dz} \;
P_{ab}^{(R)} (\as(\mu^2),z) 
\;{\widetilde f}_b({x/z},\mu^2) 
\nonumber\\ 
 & + & \sum_b 
\left\{  \int_x^{1}  
D_{ab}(\as(\mu^2))  \ 
\delta (  1 - z ) \  dz 
-  \int_0^{z_M} 
K_{ab}(\as(\mu^2))  \ 
{ 1 \over {  1 - z  }}  \ dz 
 \right\} 
  {\widetilde f}_b({x},\mu^2) \;\; .  
\end{eqnarray} 
The virtual terms in the second line 
of 
Eq.~(\ref{rewr4first}) can  be 
dealt with 
 by using the 
momentum 
sum rule, as we see next. 

\subsection{Momentum sum rule } 
\label{subsec:2d}

We will now use the momentum 
sum rule to systematically eliminate 
the $D$-terms in  
Eq.~(\ref{decompPab}) from the 
evolution equations in favor of 
the $K$-terms and $R$-terms. To 
this end, 
we  insert the  momentum 
sum rule   
\begin{equation}
\label{momsum}
 \sum_c \int_0^1  z  
 \   P_{ca}(\as(\mu^2),z) \ dz = 0 
\;\; ({\rm{for}} \;\; {\rm{any}} \;\; a)  
\;  
\end{equation} 
into the evolution equations,  
 by subtracting the 
momentum sum integral in   
 the curly bracket in the second 
line of Eq.~(\ref{rewr4first}). 
Recall from Eq.~(\ref{flavdiag}) that 
the   $D_{ab}$ and $K_{ab}$
terms in this equation are diagonal 
in flavor. Therefore, by   
interchanging  indices, we obtain 
 from Eq.~(\ref{rewr4first}) 
\begin{eqnarray}
\label{rewr5bis}
&& \frac{\partial \;{\widetilde f}_a(x,\mu^2)}{\partial  \ln \mu^2}  =  \sum_b 
\int_x^{z_M} {dz} \;
P_{ab}^{(R)} (\as(\mu^2),z) 
\;{\widetilde f}_b({x/z},\mu^2) 
\\ 
 & + &  \sum_c  
\left\{  \int_x^{1}  
D_{ca}(\as(\mu^2))  \ 
\delta (  1 - z ) \  dz 
-  \int_0^{z_M} 
K_{c a}(\as(\mu^2))  \ 
{ 1 \over {  1 - z  }}  \ dz \right. 
\nonumber\\ 
&-& \left. \int_0^1  z   
\  P_{ca}(\as(\mu^2),z) \ dz 
 \right\} 
 {\widetilde f}_a({x},\mu^2) 
 \;\; . 
\nonumber
\end{eqnarray} 

Let us  now use again the 
decomposition 
(\ref{decompPab}) for 
$ P_{ca}(\as(\mu^2),z) $ in the last 
line of Eq.~(\ref{rewr5bis}).  
We  
observe that  
 the $D_{ca}$ term 
in $ P_{ca}(\as(\mu^2),z) $ cancels 
against the first term in the 
curly bracket in Eq.~(\ref{rewr5bis}),   
while the 
$R_{ca}$ term 
in $ P_{ca}(\as(\mu^2),z) $
may be restricted to the region 
$ z < z_M$, up to order 
${\cal O}(1-z_M)$. 
Finally, the 
$K_{ca}$ term 
in $ P_{ca}(\as(\mu^2),z) $
may be combined with 
the second  term in the 
curly bracket in Eq.~(\ref{rewr5bis}). 
Putting pieces together,  we get 
\begin{eqnarray}
\label{rewr8}
&& \frac{\partial \;{\widetilde f}_a(x,\mu^2)}{\partial \ln \mu^2}  =  \sum_b 
\int_x^{z_M} {dz} \;
P_{ab}^{(R)} (\as(\mu^2),z) 
\;{\widetilde f}_b({x/z},\mu^2) 
\\ 
 & - & 
\left\{  
 \sum_c \int_0^{z_M}  z \ 
K_{ca}(\as(\mu^2)) \ 
{ 1 \over {  1 - z  }}   \ dz  
+ \sum_c \int_0^{z_M}  z 
\  R_{ca}(\as(\mu^2),z) \ dz 
 \right\} 
  {\widetilde f}_a({x},\mu^2) \;\; . 
\nonumber 
\end{eqnarray} 
We thus recognize, by using 
  Eq.~(\ref{realPab}), 
that  the evolution equations 
(\ref{rewr4first}) can be written as  
\begin{eqnarray}
\label{sudrecast}
 \frac{\partial \;{\widetilde f}_a(x,\mu^2)}{\partial \ln \mu^2}  &=&  \sum_b \left\{ 
\int_x^{z_M} {dz} \;
P_{ab}^{(R)} (\as(\mu^2),z) 
\;{\widetilde f}_b({x/z},\mu^2) \right. 
\nonumber\\ 
 & - & \left. 
  \int_0^{z_M} dz \  z 
\ P_{ba}^{(R)}(\as(\mu^2) ,  z ) \ 
 {\widetilde f}_a({x},\mu^2)
 \right\} 
  \;\; .  
\end{eqnarray}

\subsection{Sudakov form factor} 
\label{subsec:2e}

Eq.~(\ref{sudrecast}) recasts the  evolution of each parton  $a$ 
 in terms of the 
  real-emission 
probabilities $ P_{a b}^{(R)}$ and 
$ P_{b a}^{(R)}$ and of the 
resolution parameter $z_M$.   
It can be rewritten in a form 
which has the advantage 
of being solvable  by an iterative 
Monte Carlo procedure 
if we  introduce  the 
 Sudakov form factor, 
defined as   
\begin{equation}
\label{sud-def}
 \Delta_a ( z_M, \mu^2 , \mu^2_0 ) = 
\exp \left(  -  \sum_b  
\int^{\mu^2}_{\mu^2_0} 
{{d \mu^{\prime 2} } 
\over \mu^{\prime 2} } 
 \int_0^{z_M} dz \  z 
\ P_{ba}^{(R)}(\as(\mu^{\prime 2}) , 
 z ) 
\right) 
  \;\; .  
\end{equation}
The  Sudakov form factor 
$ \Delta_a ( z_M, \mu^2 , \mu^2_0 ) 
$ has the interpretation of  
probability for parton $a$ to undergo 
no branching between evolution 
scale $\mu_0$ and evolution scale 
$\mu$, where the  branchings  are  understood to be classified according 
to the given resolution $z_M$. 

Noting   that 
\begin{equation}
\label{sud-deriv}
  \frac{\partial \;{\Delta}_a(z_M,\mu^2, 
\mu^2_0)}{\partial \ln \mu^2}  
 = 
  -  \Delta_a ( z_M, \mu^2 , \mu^2_0 )  \sum_b  
 \int_0^{z_M} dz \  z 
\ P_{ba}^{(R)}(\as(\mu^2) ,  z ) 
  \;\; , 
\end{equation}
we obtain  from  Eq.~(\ref{sudrecast}) 
(removing $z_M$ and $\mu^2_0$  from the argument list 
for better readability) 
\begin{eqnarray}
\label{sudrecast1}
 \frac{\partial \;{\widetilde f}_a(x,\mu^2)}{\partial \ln \mu^2}  &=&  \sum_b 
\int_x^{z_M} {dz} \;
P_{ab}^{(R)} (\as(\mu^2),z) 
\;{\widetilde f}_b({x/z},\mu^2)  
\nonumber\\ 
 & + & 
 {1 \over 
\Delta_a (  \mu^2  ) 
} \ 
 \frac{\partial \;{\Delta}_a(\mu^2 
)}{\partial \ln \mu^2} \ 
 {\widetilde f}_a({x},\mu^2)
  \;\; .    
\end{eqnarray}
This evolution equation can be written in a  form  similar  to  Eq.~(\ref{evapp}), but now  in terms of real-emission probabilities  $P_{ab}^{(R)}$ and Sudakov form factors:   
\begin{equation}
\label{sudrecast2}
 \frac{\partial  }{\partial  \ln \mu^2} 
\left( 
{{ {\widetilde f}_a(x,\mu^2) } 
\over 
{ \Delta_a (  \mu^2  ) } } 
\right) 
 =  \sum_b 
\int_x^{z_M} {dz} \;
P_{ab}^{(R)} (\as(\mu^2),z) 
\; {{  {\widetilde f}_b({x/z},\mu^2)  }
\over { \Delta_a (  \mu^2  )  }}
  \;\; .   
\end{equation}
  Integrating this equation 
   we obtain,   
     with $ \Delta_a ( \mu^2_0) 
= 1 $, 
\begin{equation}
\label{sudintegral2}
  {\widetilde f}_a(x,\mu^2) 
 =  
\Delta_a (  \mu^2  ) \ 
 {\widetilde f}_a(x,\mu^2_0)  
+  \sum_b 
\int^{\mu^2}_{\mu^2_0} 
{{d \mu^{\prime 2} } 
\over \mu^{\prime 2} } 
{
{\Delta_a (  \mu^2  )} 
 \over 
{\Delta_a (  \mu^{\prime 2}  
 ) }
}
\int_x^{z_M} {dz} \;
P_{ab}^{(R)} (\as(\mu^{\prime 2})
,z) 
\;{\widetilde f}_b({x/z},
\mu^{\prime 2})  
  \;\;  .    
\end{equation}

We  
recognize that 
introducing 
the Sudakov form factor 
has led to an equation 
which  is an  integral equation of  Fredholm type, 
\begin{equation}
\label{fredh}
f(t) = f_0(t) + \lambda \int_a^b K(t,y) f(y)
dy     \;\;  .     
\end{equation}
This  can be solved by iteration 
 as a 
     series~\cite{Hautmann:2017xtx}   
\begin{equation}
\label{neuma} 
f(t) =
\lim_{n\to \infty} \sum_{i=0}^n \lambda^i u_i(t) \;\; , 
\end{equation}
where 
\begin{eqnarray}
u_0(t) & = &  f_0(t)  
 \;\;  , 
\nonumber \\
u_1(t) & = &  \int_a^b
K(t,y) f_0(y) dy \;\;  , \nonumber \\
u_2(t) & = & 
 \int_a^b \int_a^b
K(t,y_1)  K(y_1,y_2)f_0(y_2) 
dy_2 dy_1      \;\;  ,   \nonumber \\
\cdots  \nonumber  \\
\vdots \nonumber  \\
u_n(t) & = &  \int_a^b \int_a^b \int_a^b  
K(t,y_1) 
\cdots K(y_{n-1},y_n) f_0(y_n) dy_n \cdots dy_2 dy_1   \;\;  . 
\end{eqnarray}

We observe that while at LO in $\as$ the splitting 
functions are positive definite, this is no longer the case 
at NLO. However, although the integrands can be 
negative, the integrals over the splitting functions which appear in the evolution kernels and Sudakov form factors 
remain positive also at NLO. We will exploit this in the 
next section to apply a Monte Carlo method for solving the 
evolution equations.

\subsection{Solution of the evolution equation applying a Monte Carlo method} 
\label{subsec:2f}

The solution of the evolution equation 
can be obtained by applying a Monte Carlo method. By this method the 
problem is reduced to that of 
generating the splitting variable $z$ and the evolution scale $\mu$.

\begin{figure}[htbp]
\centering 
\includegraphics[width=0.95\textwidth, trim=90 450 5 100, clip ]{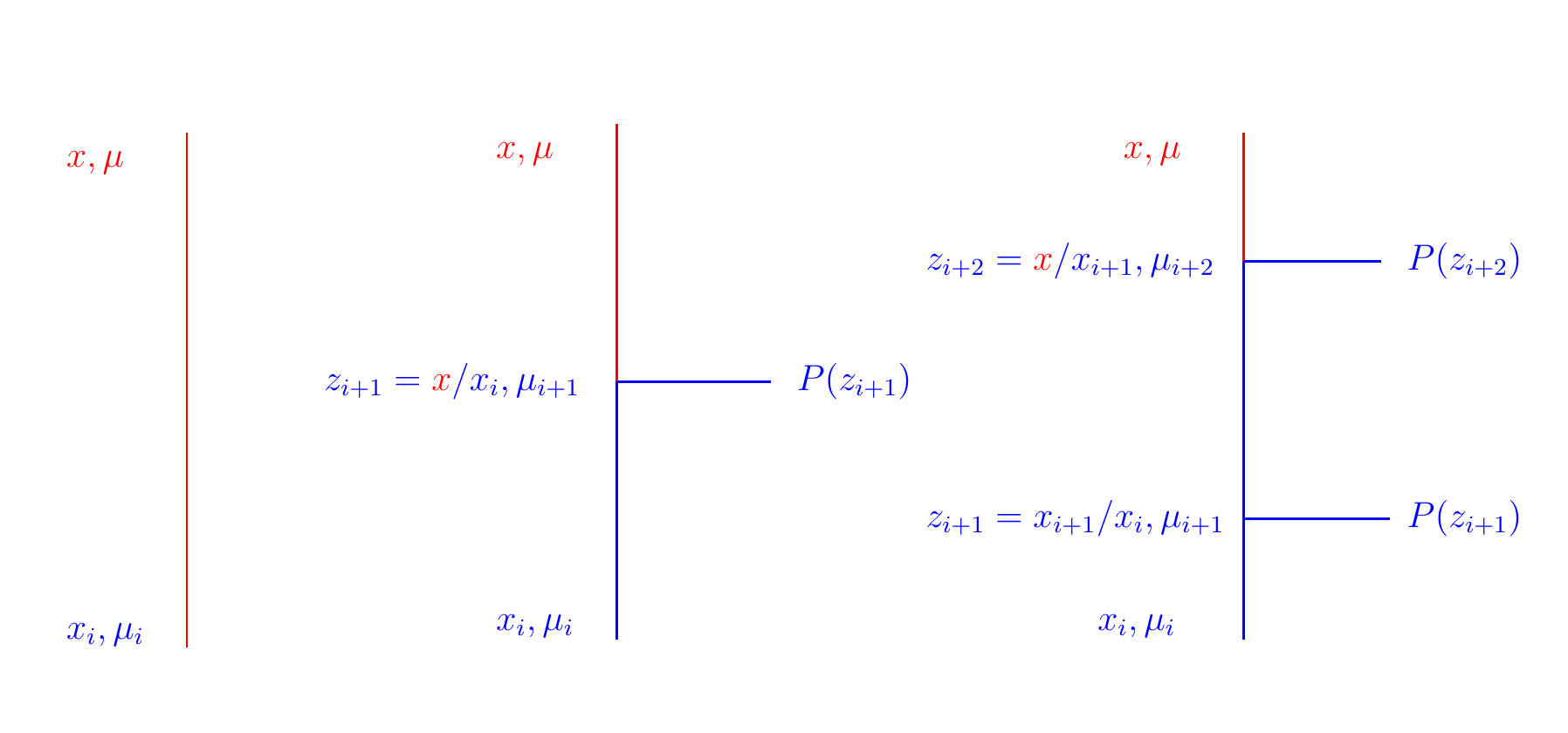}
\caption{\it   Illustration of the evolution process by iteration: a parton can evolve from scale $\mu_i$ to scale $\mu$ without  any branching (left), having one branching (middle), two branchings (right) and so on. The relevant variables are indicated.}   \label{Fig:evolution}
\end{figure}

 Fig.~\ref{Fig:evolution}   
depicts the parton evolution: we start with a parton $a$  and evolve from scale $\mu_i$ to scale $\mu$ either without any branching, or having one branching at scale $\mu_{i+1}$,  or having a second  branching at   scale $\mu_{i+2}$,  and so on.
The probability to evolve from  $\mu_{i}$ to  $\mu_{i+1}$ without any  resolvable branching is provided by  the Sudakov form factor 
$ \Delta_a ( z_M, \mu^2_{i+1} , \mu^2_i ) $. 

By introducing a random number 
$R_0$ in $[0,1]$, we  generate 
the value $\mu_{i+1}$  
by solving Eq.~(\ref{sud-def})  for $\mu_{i+1}$ at a given $\mu_{i}$,    
\begin{equation}
\label{suda-cal}
R_0 \int_{\mu_{i}^{2}}^{\mu_{max}^{2}} d \Delta_a ( z_M, \mu^{2} , \mu^2_0) =   \int_{\mu_{i}^{2}}^{\mu_{i+1}^{2}} d \Delta_a ( z_M, \mu^{ 2} , \mu^2_0) 
\; ,    
\end{equation}
that is, 
\begin{equation} 
-R_0 \left(1-\frac{\Delta_a(\mu_{max}^2)}{ \Delta_a(\mu_{i}^2)}\right) + 1  =  \frac{ \Delta_a(\mu_{i+1}^{2})}{ \Delta_a(\mu_{i}^2)} \; , 
\label{sudakov_calc}
\end{equation}
where the upper bound may be taken to be 
  $\mu_{max} \to \infty$,  leading to the simple expression 
$
R_0  = 1-  \Delta_a(\mu_{i+1}^{2})  /  \Delta_a(\mu_{i}^2) 
$.\footnote{See~\cite{Platzer:2011dq} for a detailed   
discussion of the role of integration  bounds  in the form factor.}   

The splitting variables $z$ are generated from
\begin{equation}
\int_{z_{min}}^{z_{i+1}}  dz^\prime  \ P_{ba}^{(R)}(z^\prime, \as(\mu_{i+1}) ) =  R_1 \int_{z_{min}}^{z_{M}} dz^\prime  
   \ P_{ba}^{(R)}(z^\prime,\as(\mu_{i+1})),  
\label{MC2}
\end{equation}
where  $R_1$ is a random number in $[0,1]$,  $z_{M}$ is the 
resolution parameter,  and 
 $z_{min}$ is  the lowest  
kinematically allowed value. 

Generating a pair of $z_i$, $\mu_i$ values many times,   we obtain a true and unbiased estimate of the integrals,  and a solution of the evolution equations. 

We have implemented the 
Monte Carlo method 
to solve 
 the evolution equations 
in a numerical program. 
The program is  a development of 
the code~\cite{Hautmann:2014uua}  which    was earlier employed 
by some of us 
for studies of the CCFM  
equations~\cite{hj-updfs-1312}. 
It is worth observing that 
the application    
to the case of the evolution equations 
studied in this paper 
 presents different features with respect to the 
 case of the CCFM equations. The differences   involve 
especially  the  
flavor structure of the two 
 equations, and the 
behavior of the kernels 
at small longitudinal momentum 
fractions. While CCFM equations 
are dominated by the gluon channel, 
Eq.~(\ref{sudintegral2}) has  fully coupled flavor structure. The small-$x$ 
behavior of CCFM kernels is controlled by the non-Sudakov form factor~\cite{Catani:1989sg}.  
In the case    
of Eq.~(\ref{sudintegral2}) it is essential to 
work with momentum-weighted distributions  to improve the  
convergence of the numerical  integration over the region of small $x$.  
In  Secs.~\ref{sec:3},  \ref{sec:4}, 
\ref{sec:5} we will employ 
 this program to compute numerical 
results.

\subsection{Transverse momentum distributions and ordering variables}
\label{subsec:2g}

The use  of the Sudakov form factors  and the iterative method of the previous  subsection allows one to generate step by step each resolvable branching. In addition to providing the solution of the evolution equation, this approach has the advantage of  keeping  track of  detailed information about each individual branching.  For example,  the kinematics in each branching can be calculated, similarly   to what is done in a 
parton shower process. In particular  
parton distributions can be  obtained, not only depending on $x$ and $\mu$ (as in  $ {\widetilde f}_a(x,\mu^2) $), but also depending on the transverse momentum $k_\perp$ of the propagating  parton (as in TMD parton distributions 
${\cal A}_a(x,k_\perp,\mu)$).  

\begin{figure}[htb]
\begin{center} 
\includegraphics[width=0.95\textwidth, trim=90 480 5 100, clip ]{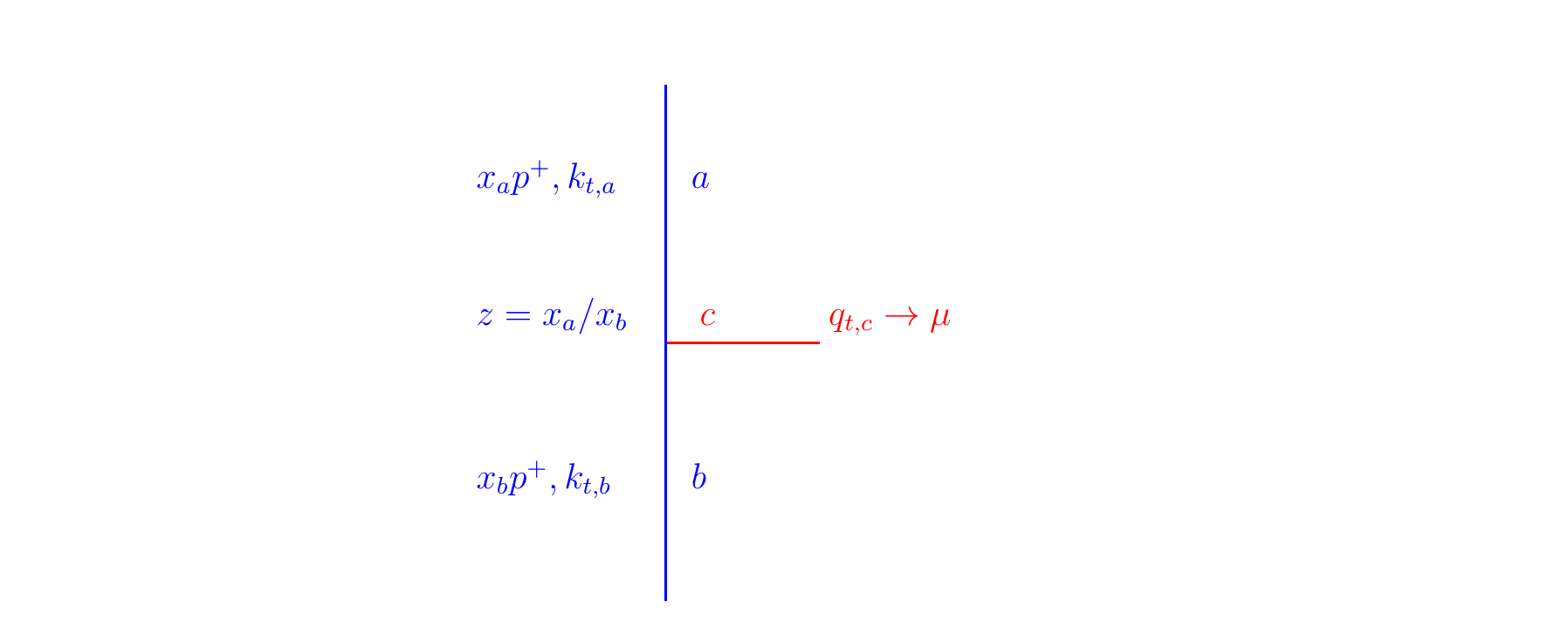}
  \caption{\it 
Branching 
process $ b \to a + c$.}
\label{fig:fig-kine}
\end{center}
\end{figure} 

Consider  the splitting  process $b \to a + c$ in  Fig.~\ref{fig:fig-kine}. 
Using the notation in the figure,  with 
   plus light-cone momenta   $p_a^+ = z p_b^+$,   $p_c^+ = (1-z) p_b^+$, we have, 
by applying conservation of 
minus light-cone momentum,   
\begin{equation}
\label{lc-cons}
p_b^2 = \frac{p_a^2 +{\bf q}_c^2}{z} + \frac{p_c^2 + {\bf q}_c^2}{1-z}  \,  , 
\end{equation} 
where ${\bf q}_c $ is the (euclidean) transverse momentum 
vector of particle $c$. 
For a space-like branching~\cite{Bengtsson:1986gz,Gieseke:2003rz,Sjostrand:2004ef} with $\mu^2 = - p_a^2$ and  $p_b^2=p_c^2=0$, 
taking  $ z \to 0$ in the 
high-energy limit       
gives 
\begin{equation} 
\label{qtord} 
\mu = 
| {\bf q}_c | \, .  
\end{equation} 
Eq.~(\ref{qtord}) is 
 referred to as   transverse momentum ordering.
If, on the other hand,  the evolution variable $\mu$ is associated with  the angle $\Theta$ 
of  the momentum of  particle $c$ with respect to the beam direction, we 
obtain the  angular ordering relation~\cite{Webber:1986mc,Catani:1990rr,Gieseke:2003rz} 
\begin{equation} 
\label{angord} 
\mu = 
| {\bf q}_c | / (1 - z) \, . 
\end{equation}

The transverse momentum  of the propagating parton 
is calculated as 
\begin{equation}
 {\bf k } = - \sum_c {\bf q}_c   \;\; .  
\label{kt_calc}
\end{equation}
The  method thus 
enables one to   determine 
the corresponding 
transverse momentum dependent (TMD) parton distribution 
$ {\cal A}_a ( x , {\bf k } , \mu^2) $,   
in addition to the inclusive 
distribution  $ {\widetilde f}_a(x,\mu^2) $, integrated over ${\bf k }$, 
\begin{equation} 
\label{unintA}
\int  
x \ {\cal A}_a ( x , {\bf k } , \mu^2)  
\  { {d^2 {\bf k }} \over \pi} 
=  {\widetilde f}_a(x,\mu^2) \; . 
\end{equation}

It has been pointed  
out in~\cite{Hautmann:2017xtx} 
that the transverse momentum 
generated 
radiatively 
by the recoils 
in the  evolution cascade depends 
strongly on the treatment of the 
non-resolvable region $z \to 1$.   
Unlike the  integrated  
distribution 
$ {\widetilde f}_a(x,\mu^2) $, 
the TMD distribution 
$ {\cal A}_a ( x , {\bf k } , \mu^2) $ 
is infrared-sensitive. The origin of this behavior lies with  singularities, present at fixed ${\bf k }$, which 
arise  from  branching 
processes in Fig.~\ref{fig:fig-kine} 
with  gluons 
at large negative rapidities,  
$ y \sim \ln q^+ / q^- \to - \infty$~\cite{Hautmann:2007uw}.   
While in the integrated 
distribution such 
singularities  cancel 
between real and virtual 
non-resolvable emissions, 
this is not, in general, 
 the case for the TMD distribution. 
As a result, supplementary conditions are needed to define the TMD  distribution  consistently~\cite{Hautmann:2007uw}.  

In the framework of the 
parton branching solution of 
evolution equations discussed in the present paper, one such set of 
conditions is provided by the    
angular ordering in 
Eq.~(\ref{angord}). 
By using the 
angular-ordered branching, 
consistent TMD distributions are 
defined, which are 
independent of the 
soft-gluon 
resolution parameter $z_M$,  for 
sufficiently large  values of $z_M$. 
In contrast,   
the transverse momentum ordered 
branching, based on   
  Eq.~(\ref{qtord}), while 
entirely suitable as long as one is 
working at the level of  integrated parton distributions, does not allow one to define 
 TMD distributions   consistently, as the transverse momentum at any given 
evolution scale would depend on the choice of the resolution parameter $z_M$. 
In~\cite{Hautmann:2017xtx} 
the above   observation is made by working  at LO  
in the strong coupling $\as$. 
In Sec.~\ref{sec:5} of the present 
 paper,   
we confirm and extend these findings  to    NLO.  

Using Eq.~(\ref{sudintegral2})   and 
 the angular ordering (\ref{angord}), we  write 
 the branching equation for 
 the evolution of TMD distributions as   
\begin{eqnarray}
\label{integeqforA}
  {\widetilde {\cal A}}_a(x,{\bf k}, \mu^2) 
 &=&  
\Delta_a (  \mu^2  ) \ 
 {\widetilde {\cal A}}_a(x,{\bf k},\mu^2_0)  
 + \sum_b 
\int
{{d^2 {\bf q}^{\prime } } 
\over {\pi {\bf q}^{\prime 2} } }
 \ 
{
{\Delta_a (  \mu^2  )} 
 \over 
{\Delta_a (  {\bf q}^{\prime 2}  
 ) }
}
\ \Theta(\mu^2-{\bf q}^{\prime 2}) \  
\Theta({\bf q}^{\prime 2} - \mu^2_0)
 \nonumber\\ 
&\times&  
\int_x^{z_M} {dz} \;
P_{ab}^{(R)} (\as({\bf q}^{\prime 2})
,z) 
\;{\widetilde {\cal A}}_b({x/z}, {\bf k}+(1-z) {\bf q}^\prime , 
{\bf q}^{\prime 2})  
  \;\;  ,     
\end{eqnarray}
where ${\widetilde {\cal A}}$ is the 
momentum weighted distribution 
${\widetilde {\cal A}} \equiv x {\cal A}$. By applying the 
method in Sec.~\ref{subsec:2f}, we solve this 
iteratively as 
\begin{equation} 
\label{itera-a} 
  {\widetilde {\cal A}}_a(x,{\bf k}, \mu^2) = \sum_{i=0}^\infty {\widetilde {\cal A}}^{(i)}_a(x,{\bf k}, \mu^2)  \;\; , 
\end{equation}
where 
\begin{equation} 
\label{itera-b} 
  {\widetilde {\cal A}}^{(0)}_a(x,{\bf k},\mu^2) 
 =   
\Delta_a (  \mu^2  ) \ 
 {\widetilde {\cal A}}_a(x,{\bf k},\mu^2_0)    \; , 
\end{equation}
\begin{eqnarray} 
\label{itera-c}  
{\widetilde {\cal A}}^{(1)}_a(x,{\bf k},\mu^2) 
& = &  \sum_b 
\int
 {{d^2 {\bf q}^{\prime } } 
\over {\pi {\bf q}^{\prime 2} } }
 \ 
{
{\Delta_a (  \mu^2  )} 
 \over 
{\Delta_a (  {\bf q}^{\prime 2}  
 ) }
}
\ \Theta(\mu^2-{\bf q}^{\prime 2}) \  
\Theta({\bf q}^{\prime 2} - \mu^2_0)
\nonumber\\ 
& \times   & 
\int_x^{z_M} {dz}  \  
P^{(R)}_{ab} (\as({\bf q}^{\prime 2})
,z) \  {\Delta_b (  {\bf q}^{\prime 2}  ) } 
\ {\widetilde {\cal A}}_b({x/z}, {\bf k}+(1-z) {\bf q}^\prime  ,  
\mu_0^{ 2})  
   \; ,  
\end{eqnarray}
and so forth. 

From the solution 
of the branching equations (\ref{sudintegral2}) 
and (\ref{integeqforA})
we will obtain  collinear and TMD  
parton distributions. The behavior at small transverse momenta, in particular, 
is controlled in this  formulation by 
the nonperturbative distributions  at scale $\mu_0$ 
and by 
 Sudakov form factors,  which 
 embody the  perturbative resummation  and are 
 defined  as  functions  
of the soft-gluon resolution scale separating resolvable and non-resolvable branchings.  
According to 
Eqs.~(\ref{sud-def}),(\ref{sudintegral2}),(\ref{integeqforA})  
this is expressed in terms of integrals 
over the  branching 
probabilities $P^{(R)} (\as ,z)$ in Eq.~(\ref{realPab}).   It depends  on the 
residues $ K_{a b}$ at the poles $z =1$  
in Eq.~(\ref{realPab}), given at one-loop and two-loop orders by the  coefficients in 
Eqs.~(\ref{oneloopK}) and (\ref{twoloopK}). 
The relationship of the $z=1$ 
behavior with 
 small  transverse momenta
is important  to construct 
reliable theoretical predictions for 
transverse momentum $ q_\perp$ spectra 
at  low $q_\perp$ in the 
production of  massive states in 
hadronic 
collisions~\cite{jcc-book,Angeles-Martinez:2015sea,ddt78,marioetalpT,pp79,css80s,johndave80s}.

We next move on to numerical results based on the 
methods described in this section. 

\section{Numerical 
parton-branching solution at NLO} 
\label{sec:3} 

In this section we present numerical 
results from the parton-branching 
solution at  NLO. 
We compare the answer thus obtained 
for the collinear 
parton density functions with the 
answer from the evolution package 
{\sc Qcdnum}~\cite{Botje:2010ay}. 
The corresponding comparison 
at LO has been shown 
in~\cite{Hautmann:2017xtx}.

For the purpose of this 
comparison
we use  as input parton 
distributions  the distributions which are the default 
set given  
in {\sc Qcdnum}. These distributions are  parameterized 
at the starting scale $\mu_0^2 = 2$~GeV$^2$ 
as 
\begin{eqnarray}
xu_v(x)  &=   &5.11  \; x^{0.8} (1-x)^3, \;\;\;  \;\;\;\;\;
 x\bar{u}(x)  =   0.19 \; x^{-0.1} (1-x)^7  , \nonumber \\
xd_v(x)  &=  &3.06 \; x^{0.8}  (1-x)^4, \;\;\;  \;\;\;\;\;
 x\bar{d}(x)  =  0.19 \; x^{-0.1} (1-x)^6 , \nonumber \\
x\bar{s}(x)   & = &  0.2 \; (x\bar{d}(x) +x\bar{u}(x) ), \;\;\;   xs(x)  =  x\bar{s}(x) , \nonumber \\
xg(x)  &= & 1.7 \; x^{-0.1}   (1-x)^5. \;\;\;   
\end{eqnarray}
Three light active flavors are assumed at the starting scale, while 
charm and bottom quarks are produced during the evolution,  for 
evolution  scales $\mu > m_c=1.73$ GeV and $\mu >  m_b= 5.0$ GeV  respectively. 
The running  coupling $\alpha_s$ is used at two loops,  with 
 $\alpha_s(m_Z^2) = 0.118$.

\begin{figure}[htbp]
\begin{center}
\includegraphics[width=7.5cm]{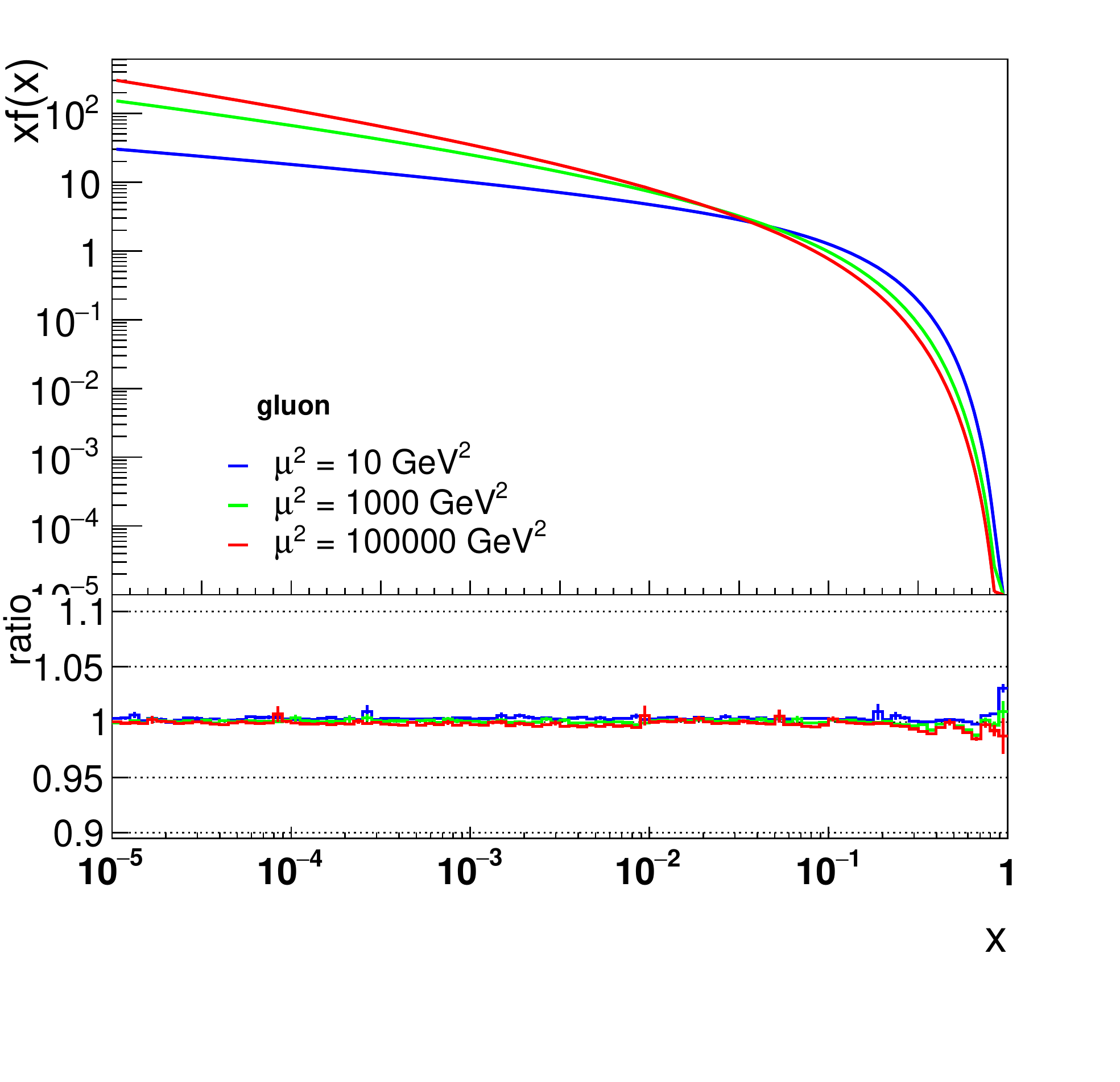}\includegraphics[width=7.5cm]{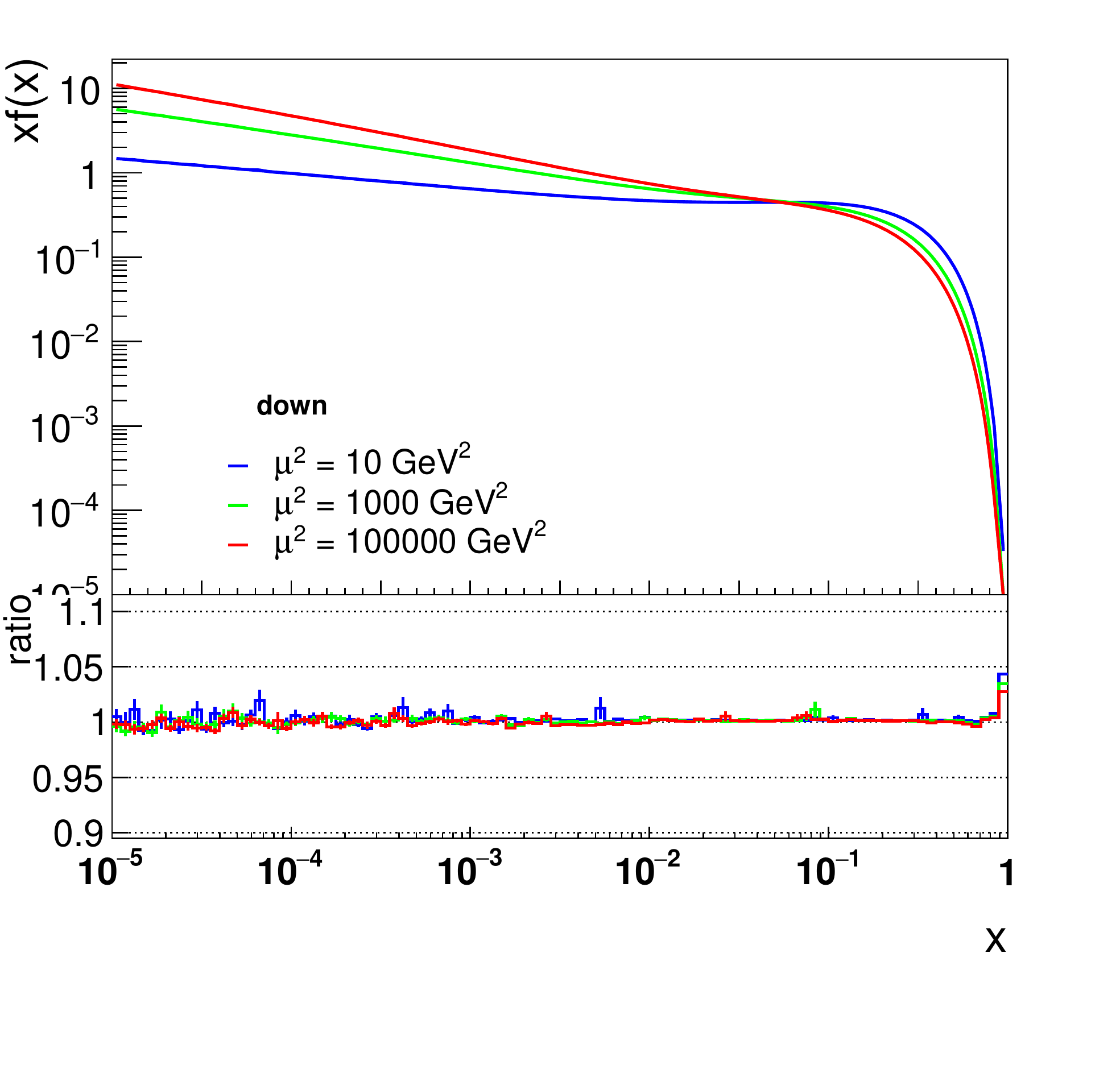}
\includegraphics[width=7.5cm]{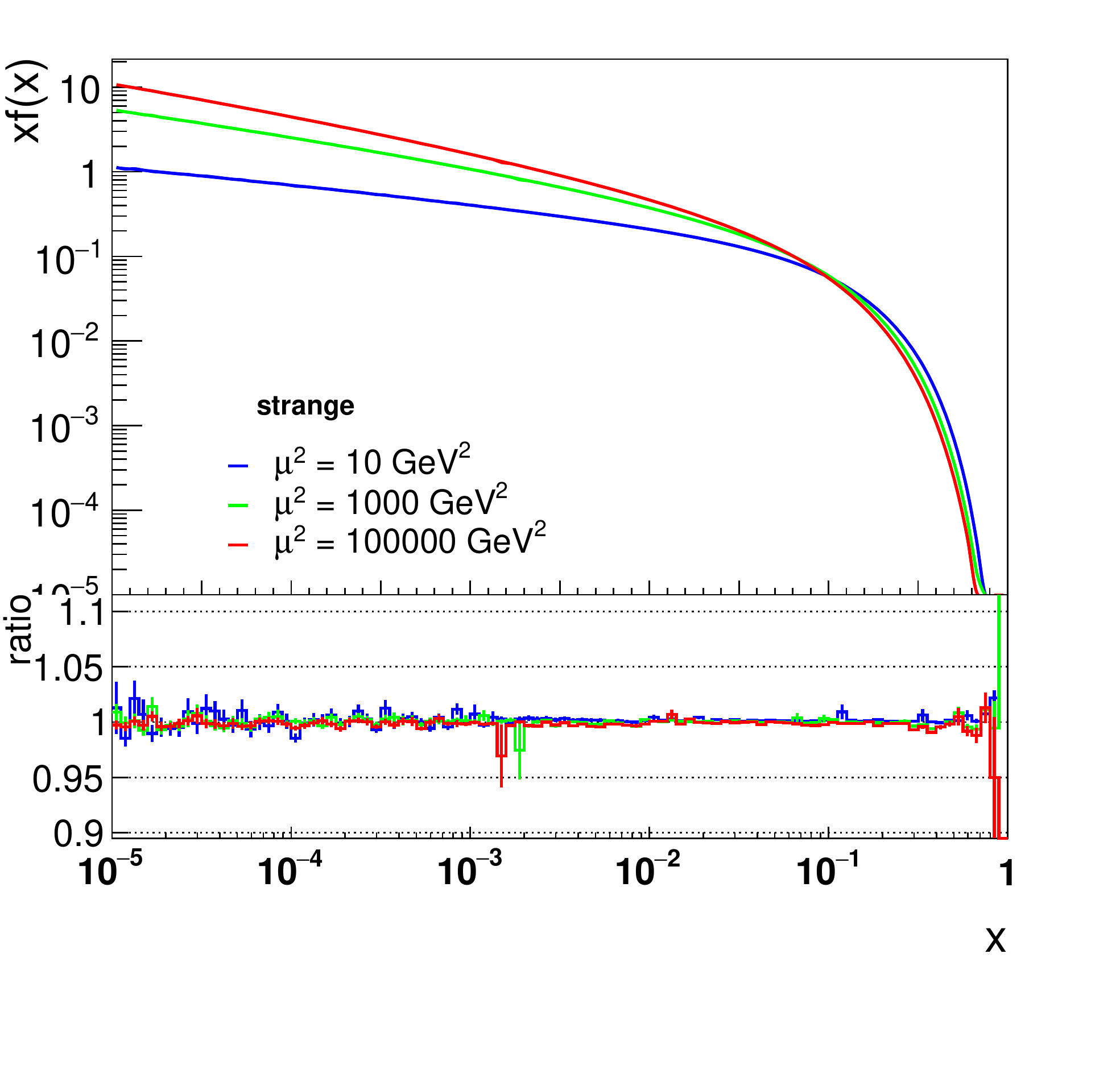}\includegraphics[width=7.5cm]{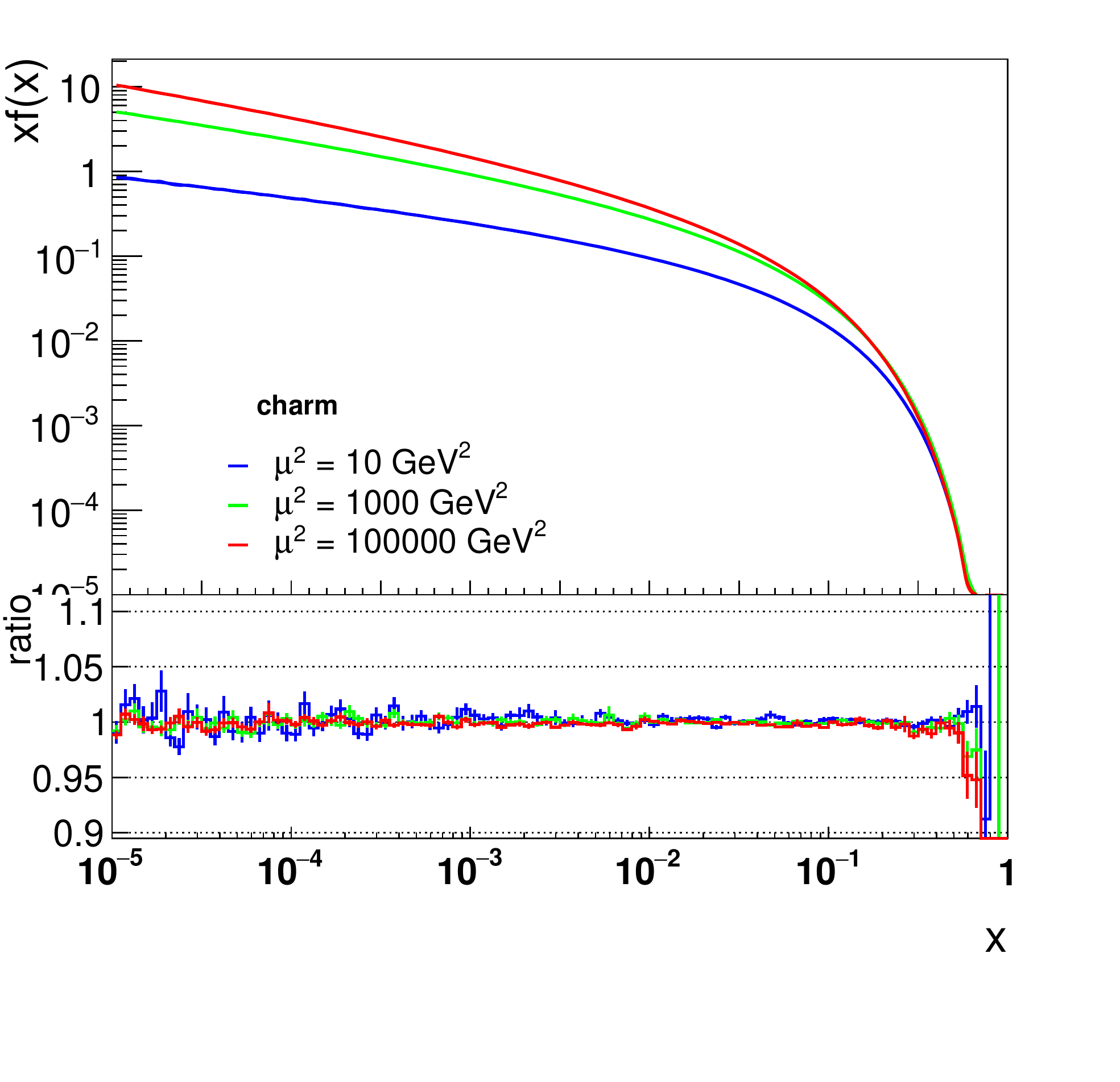}
\caption{\it  The transverse momentum integrated 
(iTMD) 
parton densities obtained from the parton branching solution,  compared with the prediction from {\sc Qcdnum}. The densities are evolved up to different scales $\mu^2$ using splitting kernels at NLO. The ratio plots show the ratio of the curves obtained with the parton branching method divided by the prediction from {\sc Qcdnum}.}
\label{fig:fig3}
\end{center}
\end{figure}

In Fig.~\ref{fig:fig3} we show the 
momentum-weighted parton densities 
(integrated over transverse momenta, iTMD) 
obtained from the parton branching solution of the evolution equations,   compared to the predictions obtained 
by  {\sc Qcdnum},  
starting from $\mu_0^2=2$~GeV$^2$
for different scales $\mu^2 = 10,\;10^3, \;10^5$~GeV$^2$. 
The parton branching solution is obtained for a fixed value of $z_{M} = 1-10^{-5}$.\footnote{The resolution scale parameter $z_M$ is in general a function 
of $\mu$. For   numerical 
illustrations in this paper we limit 
ourselves to 
taking fixed values of $z_M$.}
The overall agreement between the parton-branching  and 
{\sc Qcdnum}  calculations is better than 
 1\%  throughout the whole range in $x$ and $\mu^2$. 


\begin{figure}[htb]
\begin{center} 
\includegraphics[width=7.5cm]{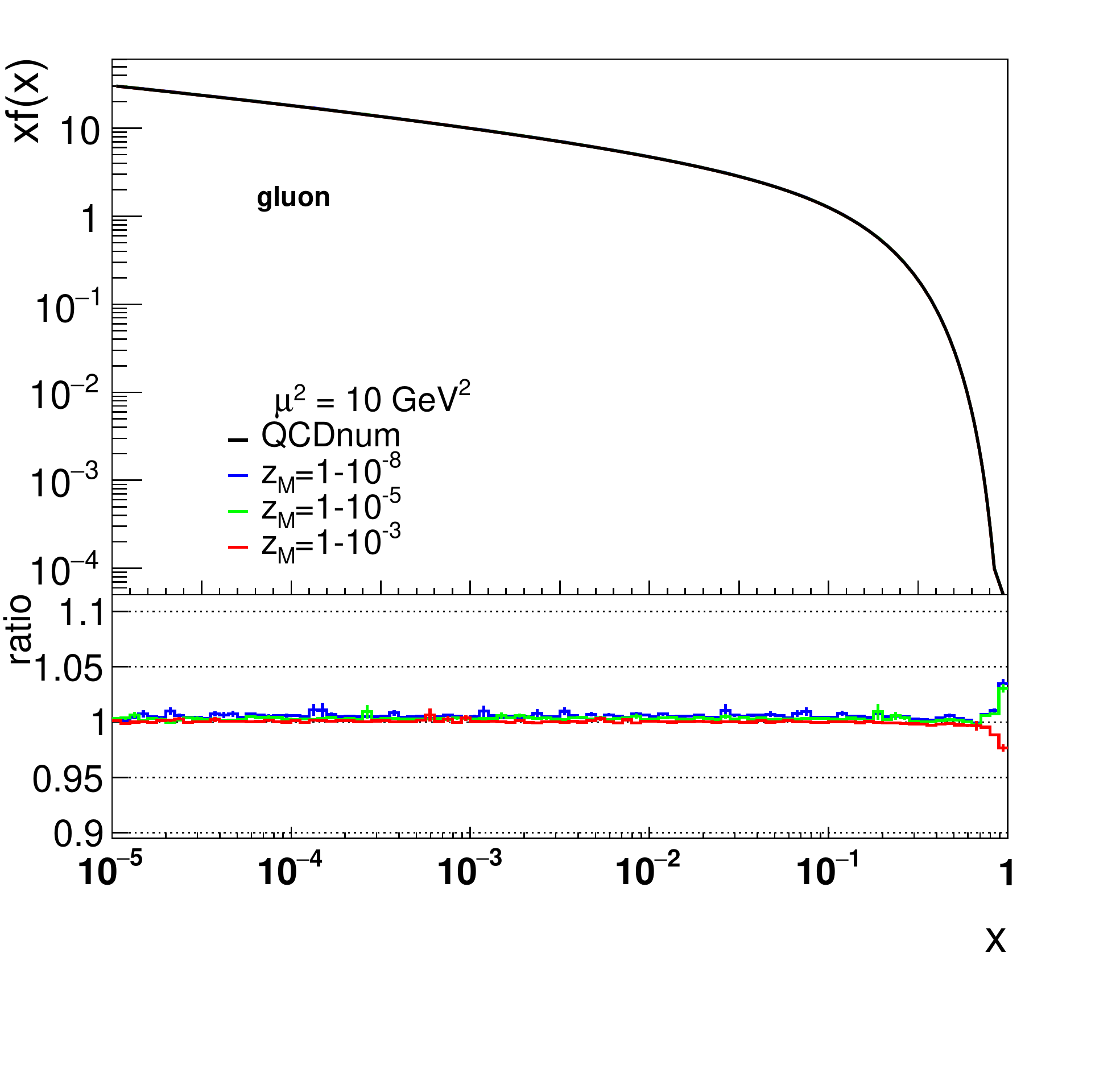}\includegraphics[width=7.5cm]{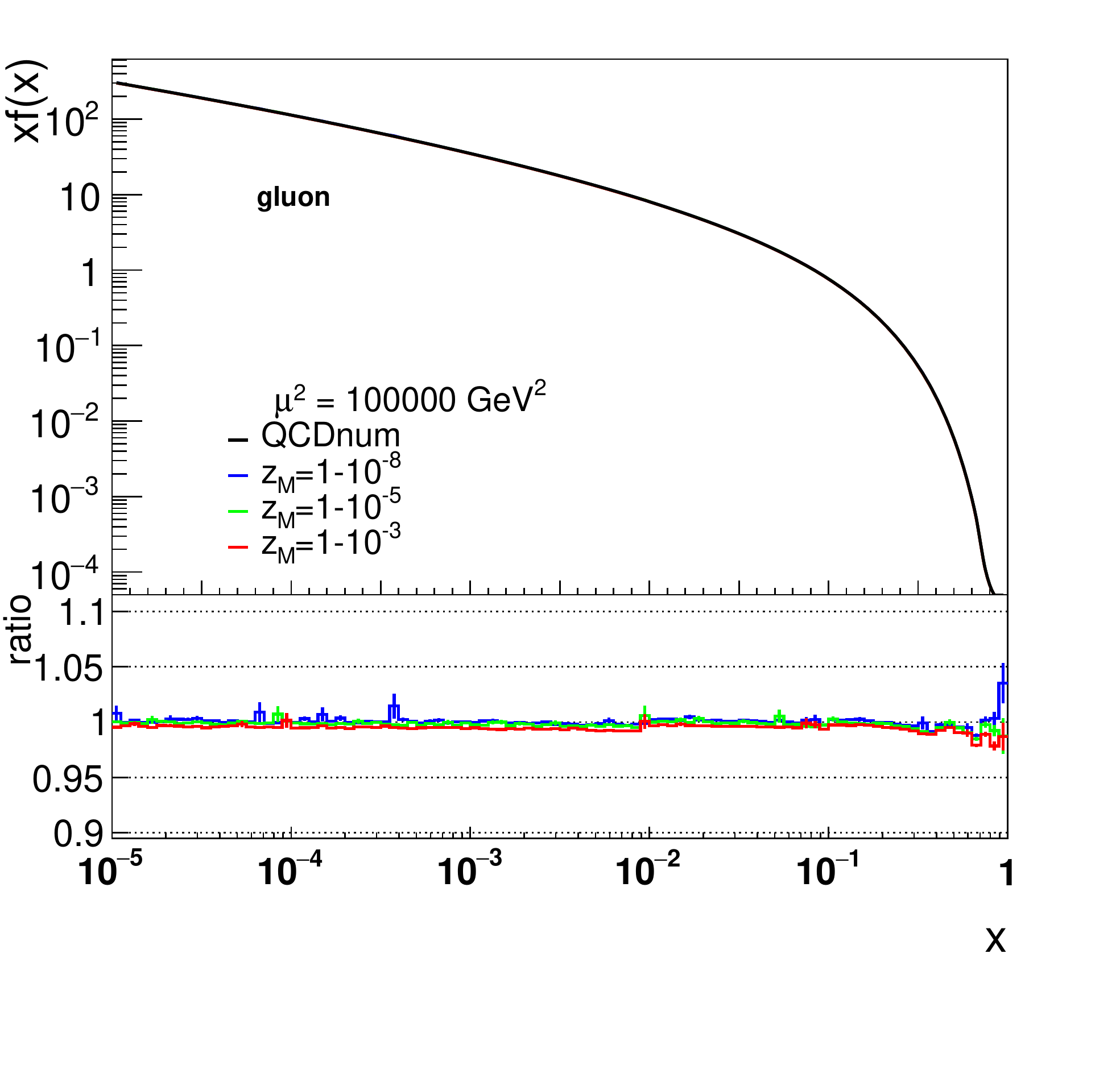}
\includegraphics[width=7.5cm]{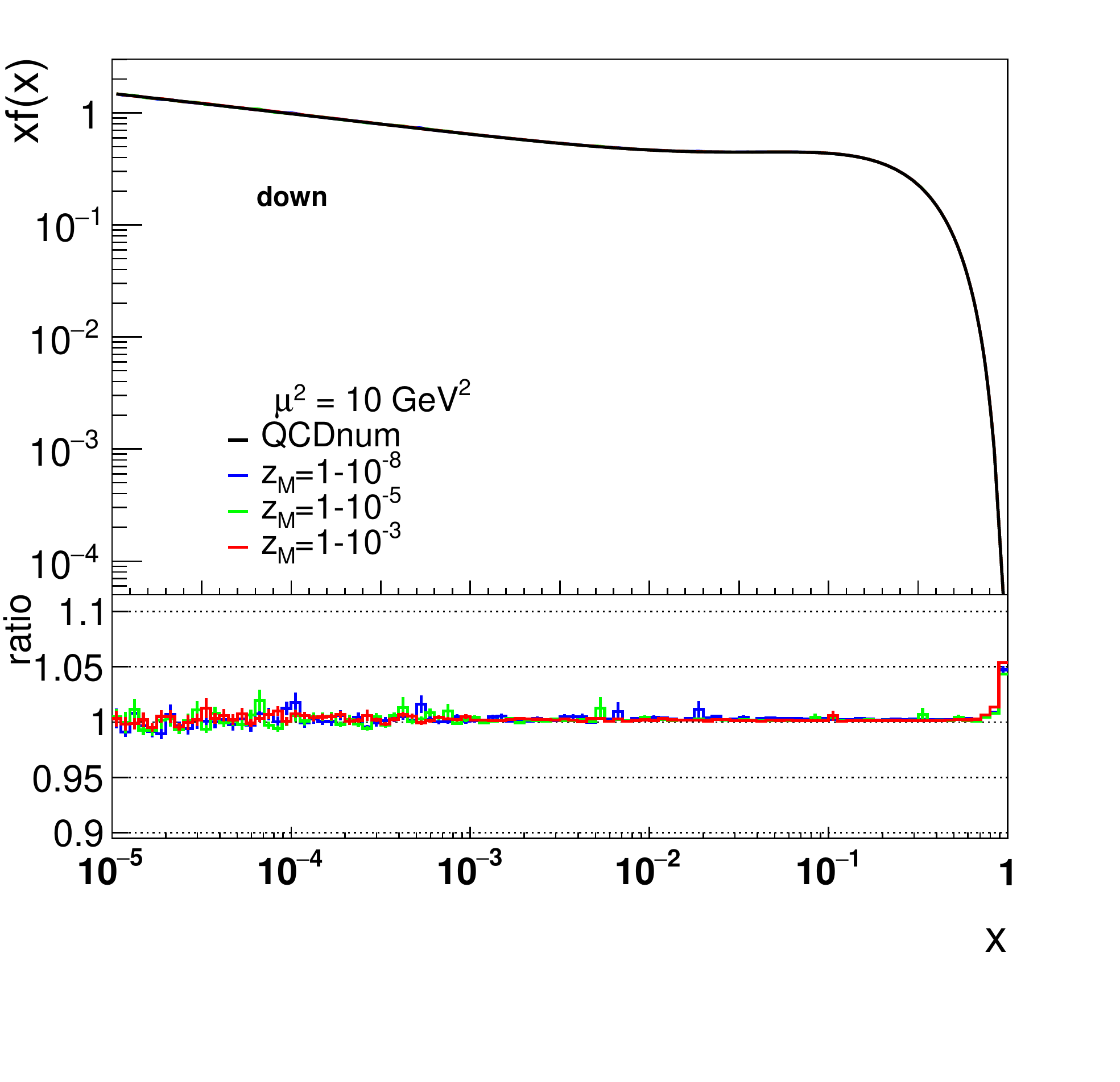}\includegraphics[width=7.5cm]{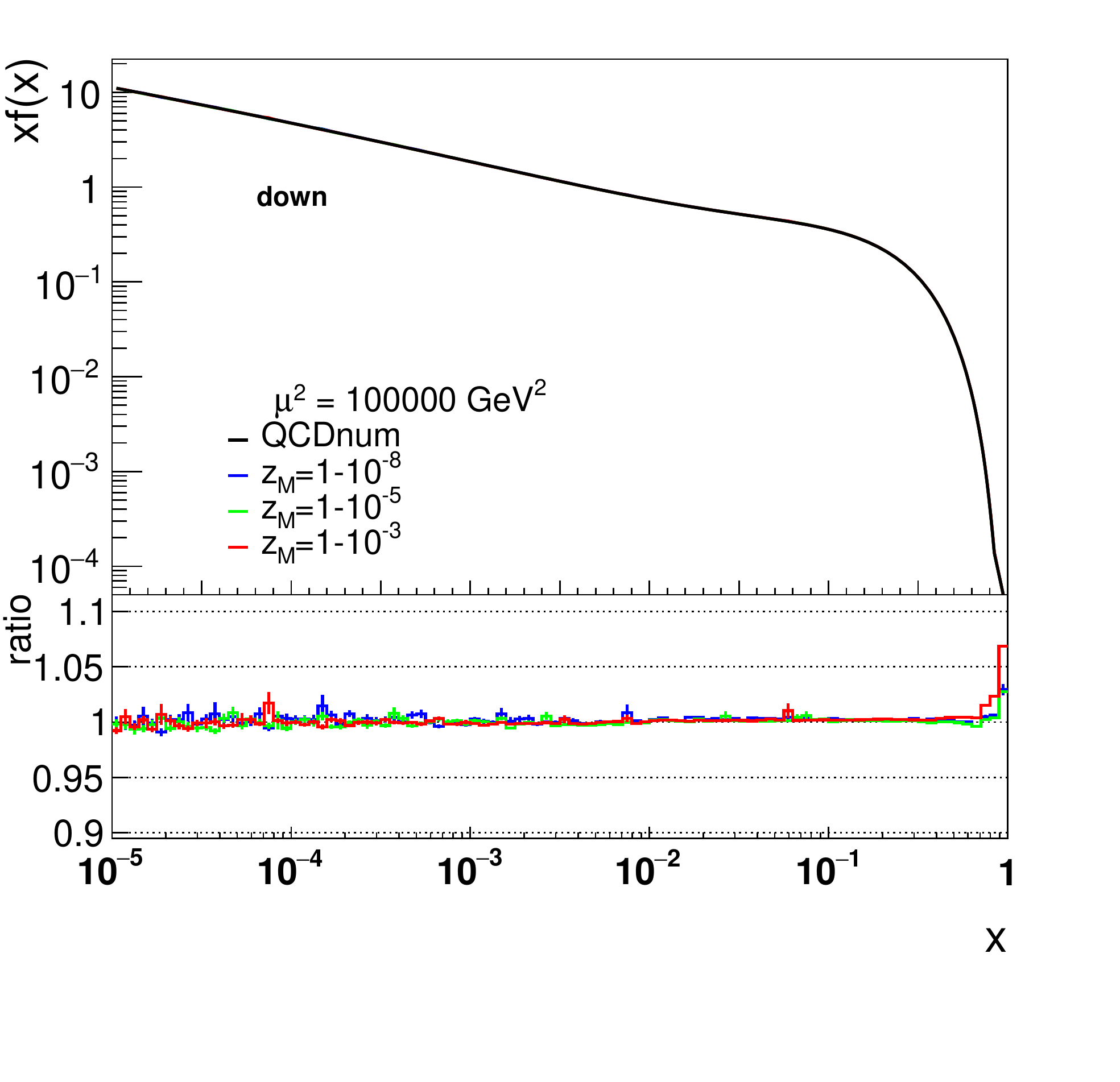}
\caption{\it  The transverse momentum 
integrated (iTMD)   parton 
densities obtained from the parton branching solution,   
compared 
with the prediction from {\sc Qcdnum}. 
The densities are  evolved up to different scales $\mu^2$ using splitting kernels at NLO, for different values of $z_M$. The ratio plots show the ratio of the curves obtained with the parton branching method divided by the prediction from 
{\sc Qcdnum}.}
\label{fig:fig4}
\end{center}
\end{figure}

In Fig.~\ref{fig:fig4} we show a comparison using different values 
of the soft-gluon resolution scale  parameter $z_{M}$. 
For all $z_M$ values chosen, no dependence on the actual choice of the  parameter $z_M$ is observed,  confirming the findings 
of~\cite{Hautmann:2017xtx}, and extending them to  NLO  in accord 
with the formalism 
 developed  in  Sec.~\ref{sec:2}. 


\section{Fit to precision DIS data} 
\label{sec:4} 

The initial parton density distributions have to be  determined from  fits  to experimental data. 
A general tool to perform such  fits  to collider  
measurements 
 is the  \verb+xFitter+ package \cite{Alekhin:2014irh}. 
 As an application of our formalism, 
 in  this  section  we  describe the  method 
 and  results of a 
  fit to the precision DIS data~\cite{Abramowicz:2015mha}. 

To start with, we implement 
 the parton branching solution of the evolution equation  in the \verb+xFitter+ package \cite{Alekhin:2014irh}.  However, using  a full Monte Carlo solution of the evolution equation for every new set of initial parameters would be too time-consuming to be efficient. Instead, we 
employ the approach developed 
in~\cite{hj-updfs-1312,Hautmann:2014uua}. Following this 
approach, first a kernel $ {\cal \widehat A}^b_{a}\left(x^{\prime \prime } ,\mu^2\right) $ is determined from the Monte Carlo  solution of the evolution equation for any initial parton of flavor $b$ evolving to a final parton of flavor $a$\footnote{In practice, since the initial state partons can be only light quarks or gluons, it is enough to determine the kernel $ {\cal \widehat A}^b_{a}$  only for one  initial  state quark and a gluon.}; then, 
this  is folded with the 
non-perturbative starting 
distribution ${\cal A}_{0,b} (x)$:  
\begin{eqnarray}
{\widetilde f}_a(x,\mu^2) 
 &= &x\int dx^\prime \int 
dx^{\prime \prime } {\cal A}_{0,b} 
(x^\prime) {\cal \widehat A}^b_a\left(
x^{\prime \prime },\mu^2\right) 
 \delta(x^{\prime  } 
x^{\prime \prime } - x) 
\nonumber  
\\
& = & \int dx^\prime 
{{\cal A}_{0,b} (x^\prime) }  
\cdot \frac{x}{x^\prime} \ { {\cal \widehat A}^b_{a}\left(\frac{x}{x^\prime},\mu^2\right) }  \;\; .
\label{iTMD_kernel}
\end{eqnarray}
The kernel  ${\cal \widehat A}^b_{a}$ includes the full parton evolution as in Eq.~(\ref{sudintegral2}), with  Sudakov form factors and splitting functions,  and  is determined with the parton branching method described earlier.  
The kernel ${\cal \widehat A}$ can be determined as a function of $x,\mu$ for the $k_\perp$-integrated iTMD distributions, or depending on 
$x,k_\perp,\mu$ for the transverse momentum dependent (TMD) distributions. 
The kernel is then folded with the initial condition ${{\cal A}_{0,b} (x^\prime) }  $. 
\begin{figure}[htb]
\begin{center}
\includegraphics[width=5.3cm]{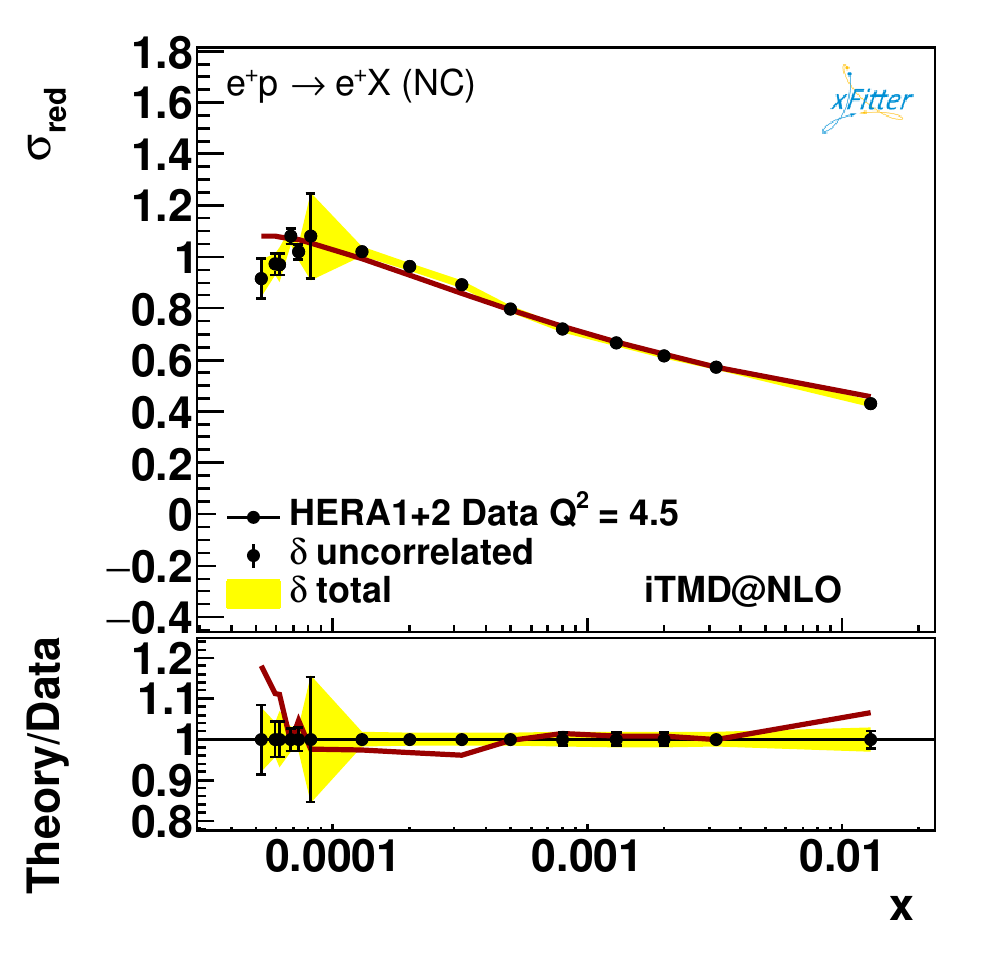}\includegraphics[width=5.3cm]{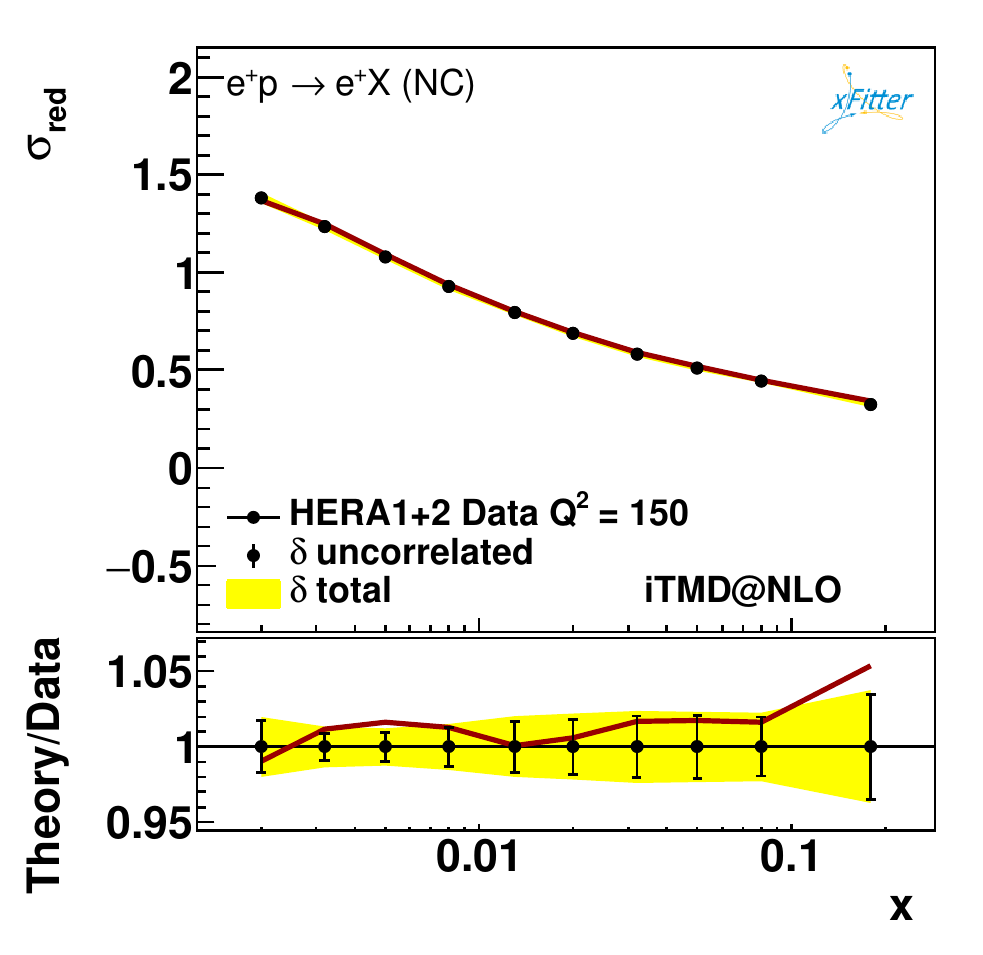}\includegraphics[width=5.3cm]{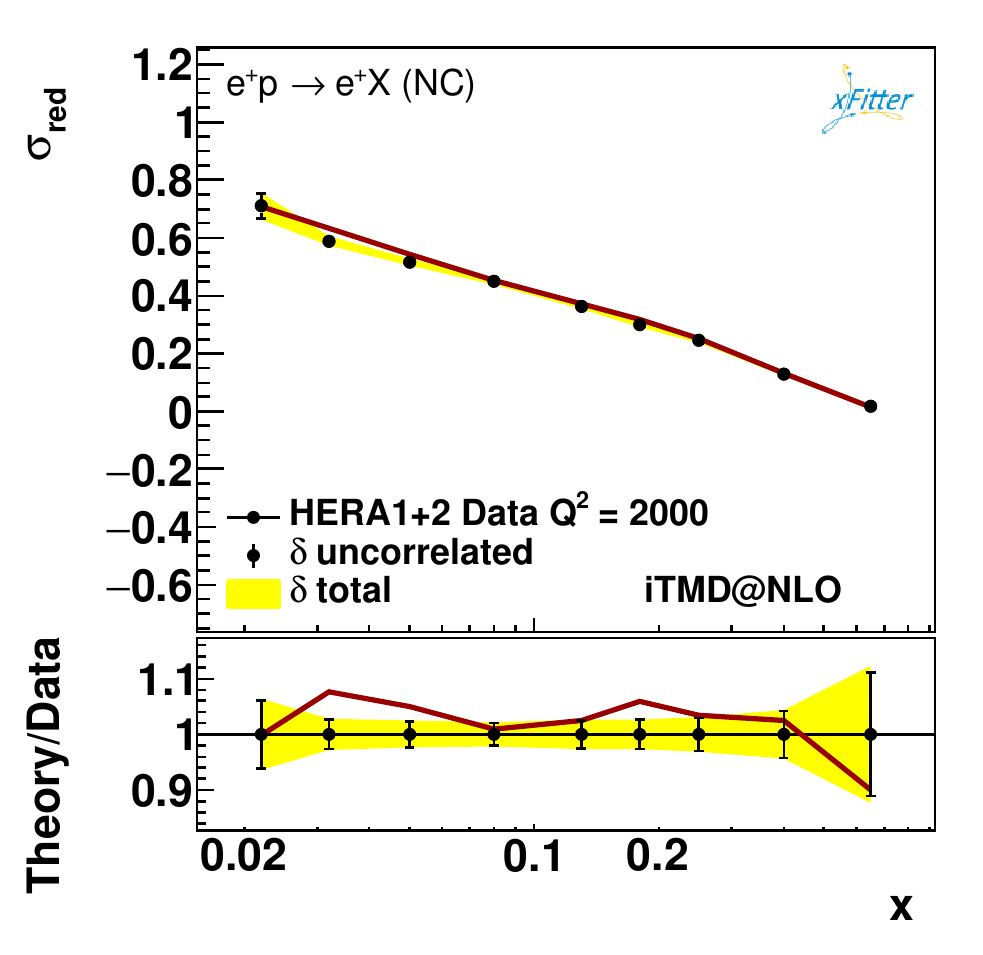}
\caption{\it  The reduced cross section $\sigma_{red}$ as measured at HERA compared to the NLO  fits from  the parton branching method (red line) for different values of $Q^2$, obtained using xFitter.}
\label{fig:fig5}
\end{center}
\end{figure}
The integrated parton density Eq.~(\ref{iTMD_kernel}) can be then used within the \verb+xFitter+ package to calculate the 
cross sections 
and to determine the parameters of the starting distributions ${\cal A}_{0,b} (x)$. 
We use precision measurements, in neutral-current  and charged-current interactions at various beam energies from HERA 1+2 \cite{Abramowicz:2015mha},  of 
the reduced cross section 
\begin{equation}
\label{eqforfits} 
\sigma_{red} = \frac{d^2 \sigma^{ep}}{dx dQ^2} \cdot \frac{Q^4 x}{2 \pi \alpha^2 (1 + (1-y)^2)}
\end{equation} 
in the  range $3.5< Q^2 < 30000$~GeV$^2$. 

Using 
two-loop running coupling with   $\alpha_s(m_Z) = 0.118$, 
  starting scale for evolution  $\mu_0^2 = 2$~GeV$^2$, heavy-quark masses   $m_c=1.73 $~GeV, $m_b= 5.0$~GeV and $m_t=175$~GeV,  and  a fixed $z_{M}=1-10^{-5}$, a very good $\chi^2/ndf \sim  1.2$ for 1132 $ndf$ is obtained for $3.5< Q^2 < 30000$~GeV$^2$. 

In Fig.~\ref{fig:fig5} the calculated 
cross section 
 $\sigma_{red}$, obtained from the fit using \verb+xFitter+, is compared with  the precision measurements from HERA \cite{Abramowicz:2015mha} for different values of $Q^2$, showing  very good agreement from low to high values of $Q^2$.

Comparing this result with the 
fit~\cite{hj-updfs-1312} to precision 
DIS data based on CCFM evolution 
equations, note that in the 
case~\cite{hj-updfs-1312} 
the constraint $x <  $  0.005 is 
applied on the data set, while 
no $ x $ constraint is applied in 
the  present case. 
By 
the 
approach of this paper the description of precision DIS measurements 
can be significantly extended toward higher $x$ and thus higher $Q^2$, 
without on the other hand 
introducing any extra constraint 
cutting out  lower $x$ data. 

We plan to analyze fits to 
data further in a future work. 

\section{TMD  densities} 
\label{sec:5} 

By applying the parton branching method of this paper, we are able to construct TMD parton densities as described in 
Sec.~\ref{subsec:2g}. 
 While large transverse momenta are generated  by perturbative evolution, the nonperturbative region of small 
 $k_\perp$ cannot be predicted in our approach but is parameterized by 
nonperturbative distributions  
 which are to be 
determined from experimental 
measurements. For the  
calculations of this section 
we use the   
parameterizations 
given in Sec.~\ref{sec:3} and 
 take simple 
  gaussian  distributions $\exp(-| k_\perp^2 | / \sigma^2)$. 
  The widths $\sigma$ are in general flavor-dependent. 
  For the numerical illustrations that follow we  
  take the same width for all parton species,  $ \sigma^2  =  q_0^2 / 2 $  for all flavors with  $q_0 = 0.5$~GeV. 
\begin{figure}[htb]
\begin{center} 
\includegraphics[width=7cm, trim=90 150 5 180, clip ]{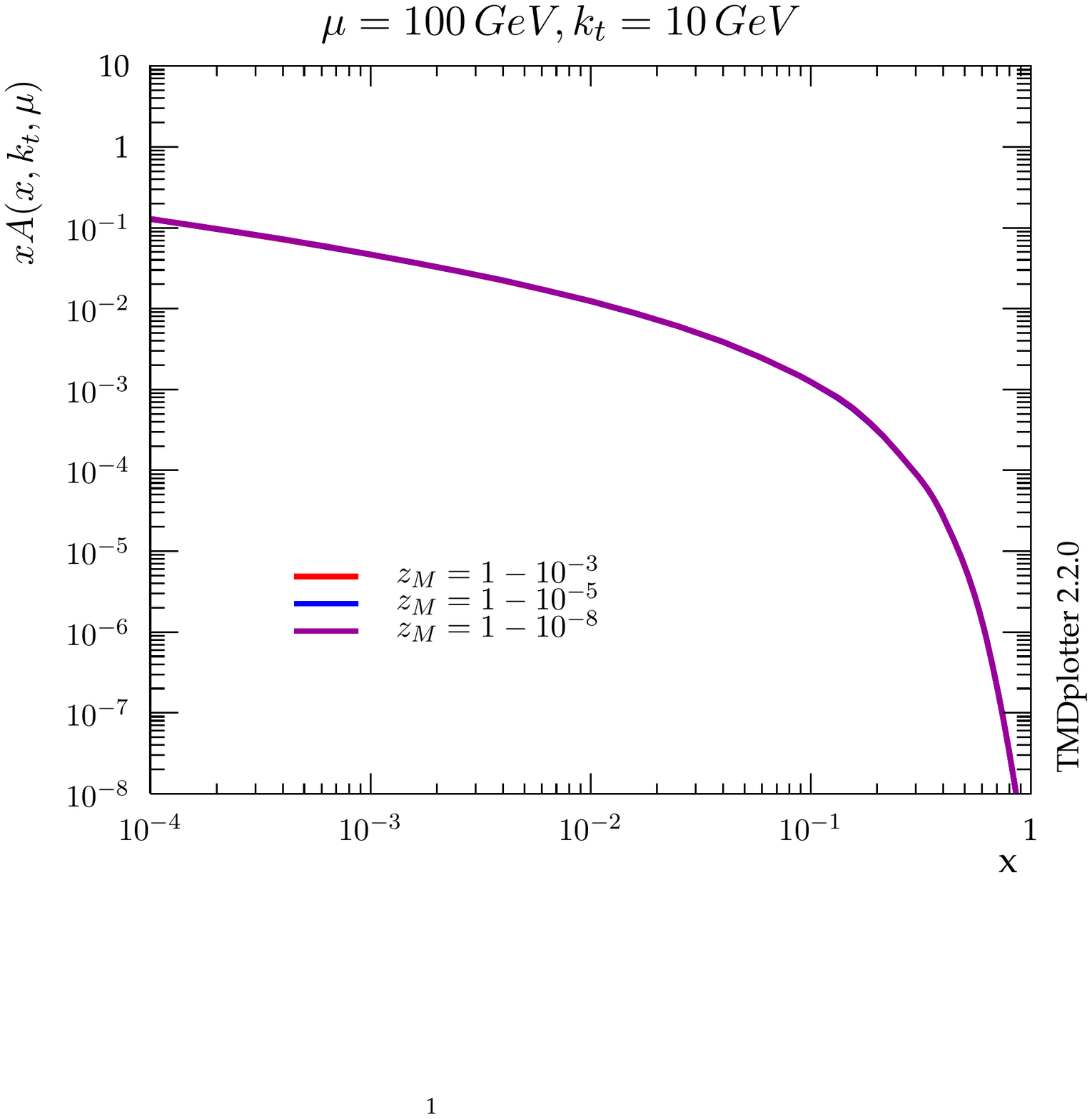} \hspace*{ 1.0cm}
\includegraphics[width=7cm, trim=90 150 5 180, clip ]{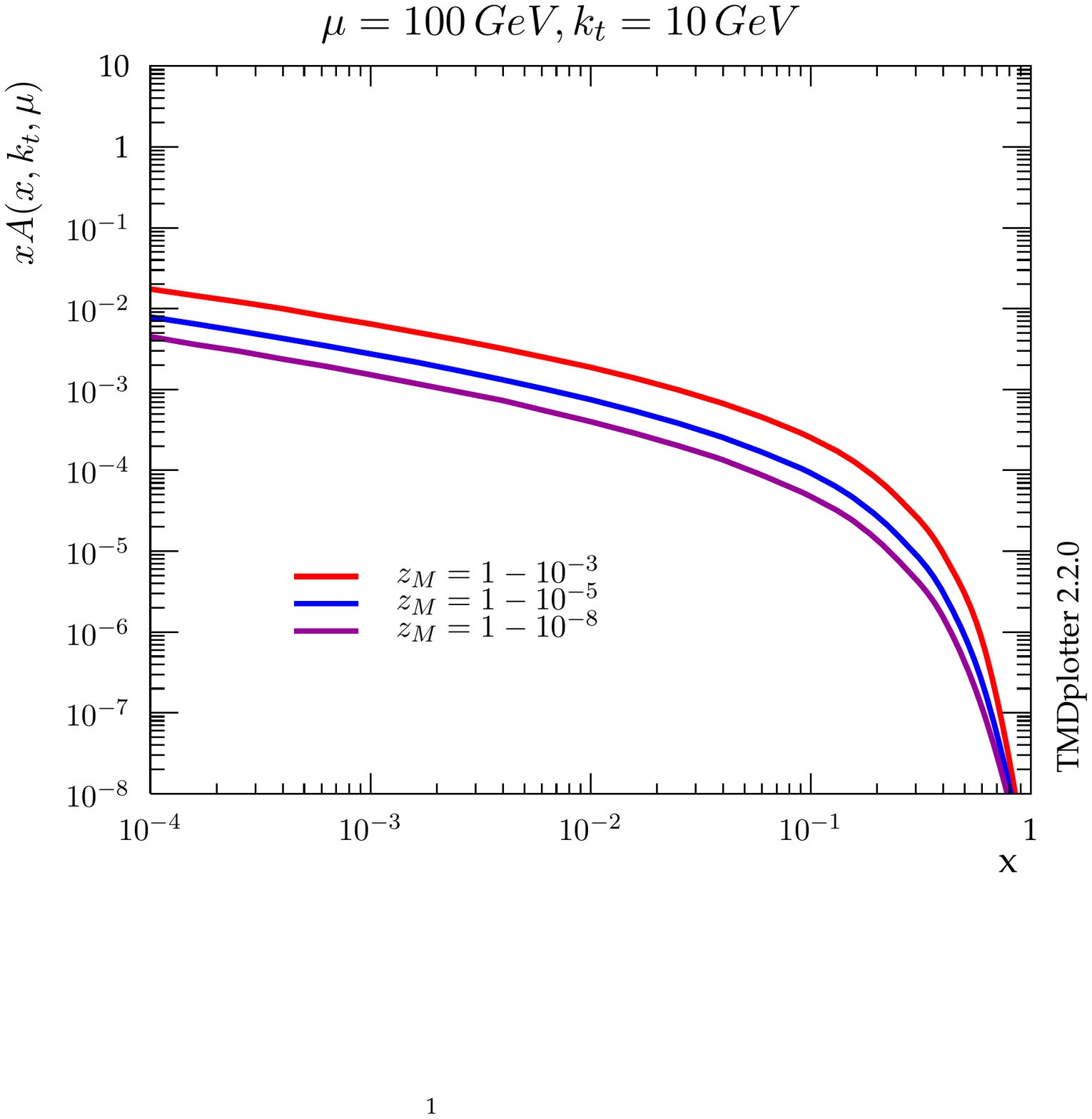}
\includegraphics[width=7cm, trim=90 150 5 180, clip ]{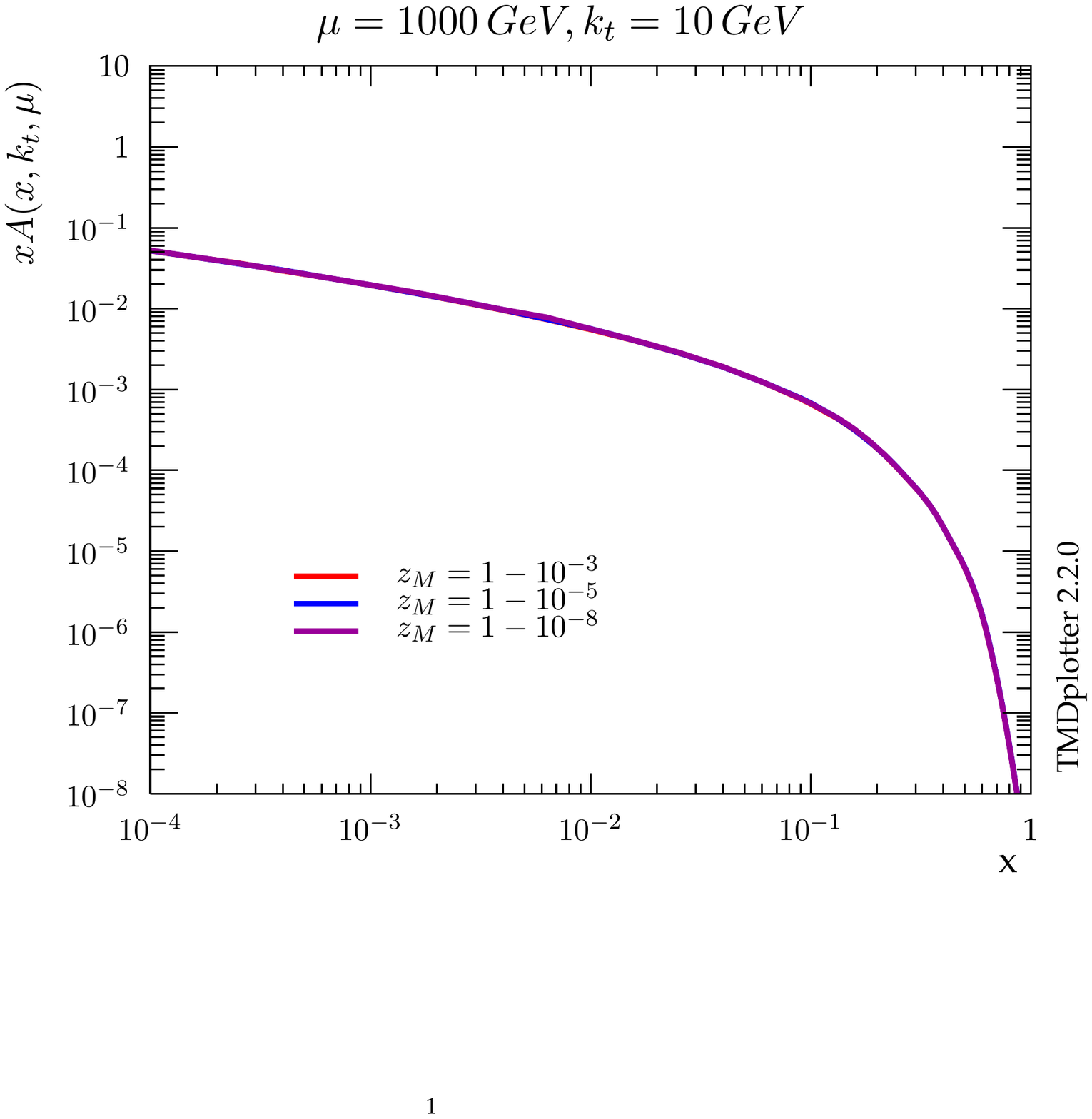} \hspace*{ 1.0cm}
\includegraphics[width=7cm, trim=90 150 5 180, clip ]{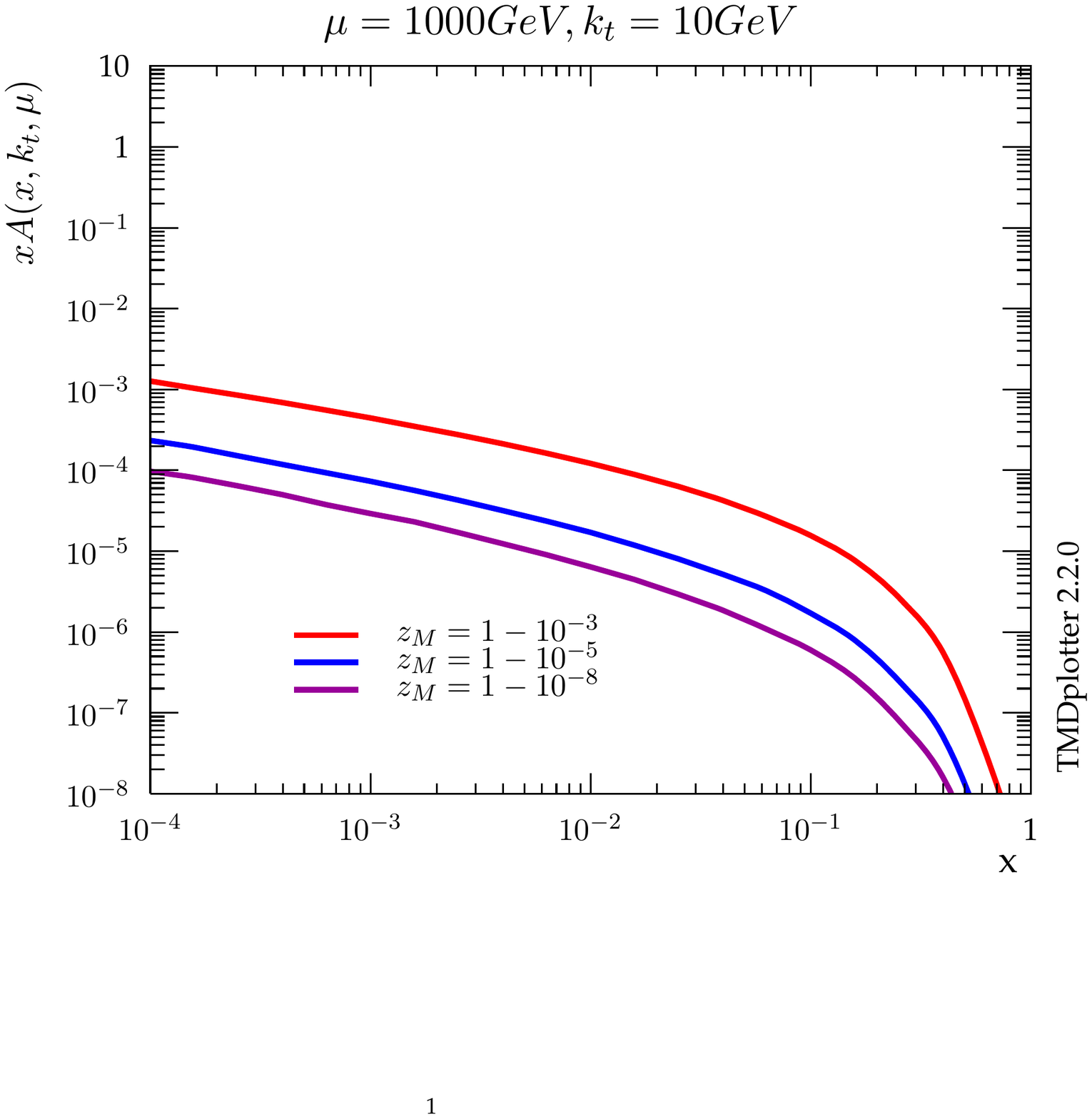}
\caption{\it  Transverse momentum  
gluon  distribution versus $x$ at $\kt = 10 \ {\rm{GeV}}  $
and  $\mu=100 \ {\rm{GeV}}  $  
(upper row), $\mu=1000 \ {\rm{GeV}}  $   
(lower row)
for different values of the resolution 
scale 
parameter  $z_{M}=1-10^{-3}, 1-10^{-5}, 1-10^{-8}$:  (left) 
angular ordering;  (right)  transverse momentum ordering.}
\label{fig:fig6}
\end{center}
\end{figure}
We present numerical 
results for the evolution of 
TMD parton densities, 
using  the parton branching 
solution of the evolution equations 
with NLO evolution kernels.

\begin{figure}[htb]
\begin{center} 
\includegraphics[width=7cm, trim=90 150 5 180, clip ]{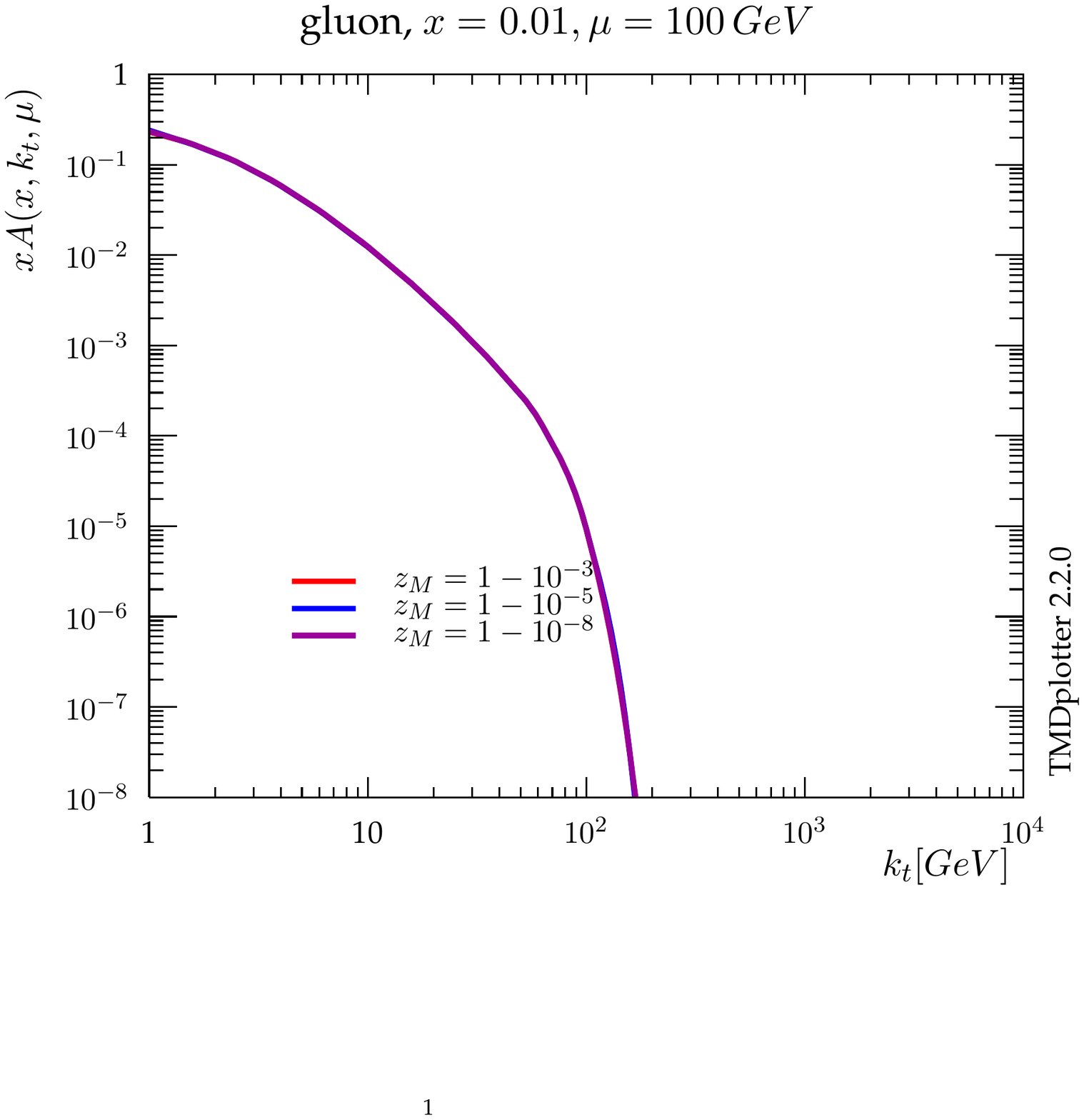} \hspace*{ 1.0cm}
\includegraphics[width=7cm, trim=90 150 5 180, clip ]{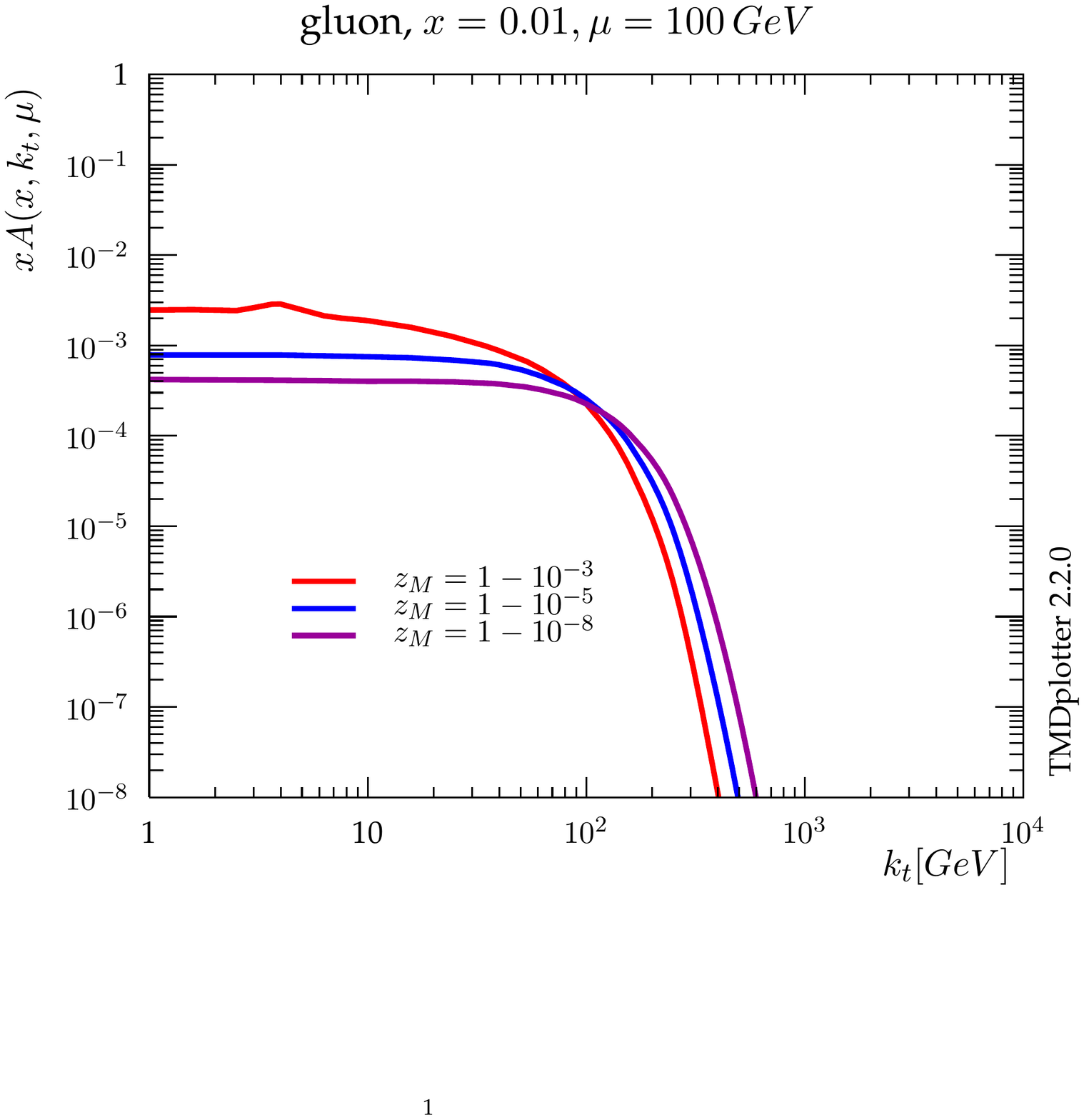}
\includegraphics[width=7cm, trim=90 150 5 180, clip ]{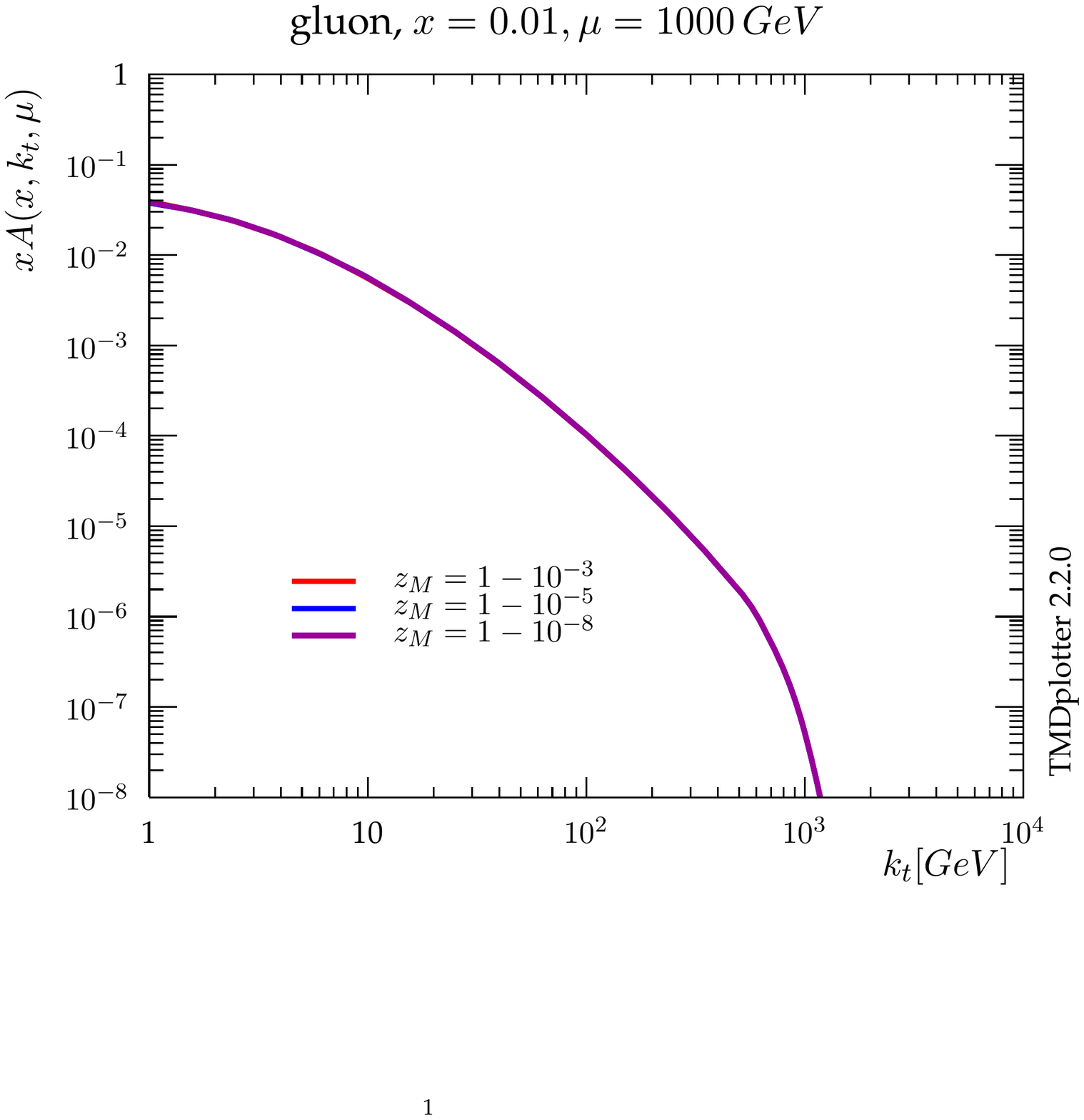} \hspace*{ 1.0cm}
\includegraphics[width=7cm, trim=90 150 5 180, clip ]{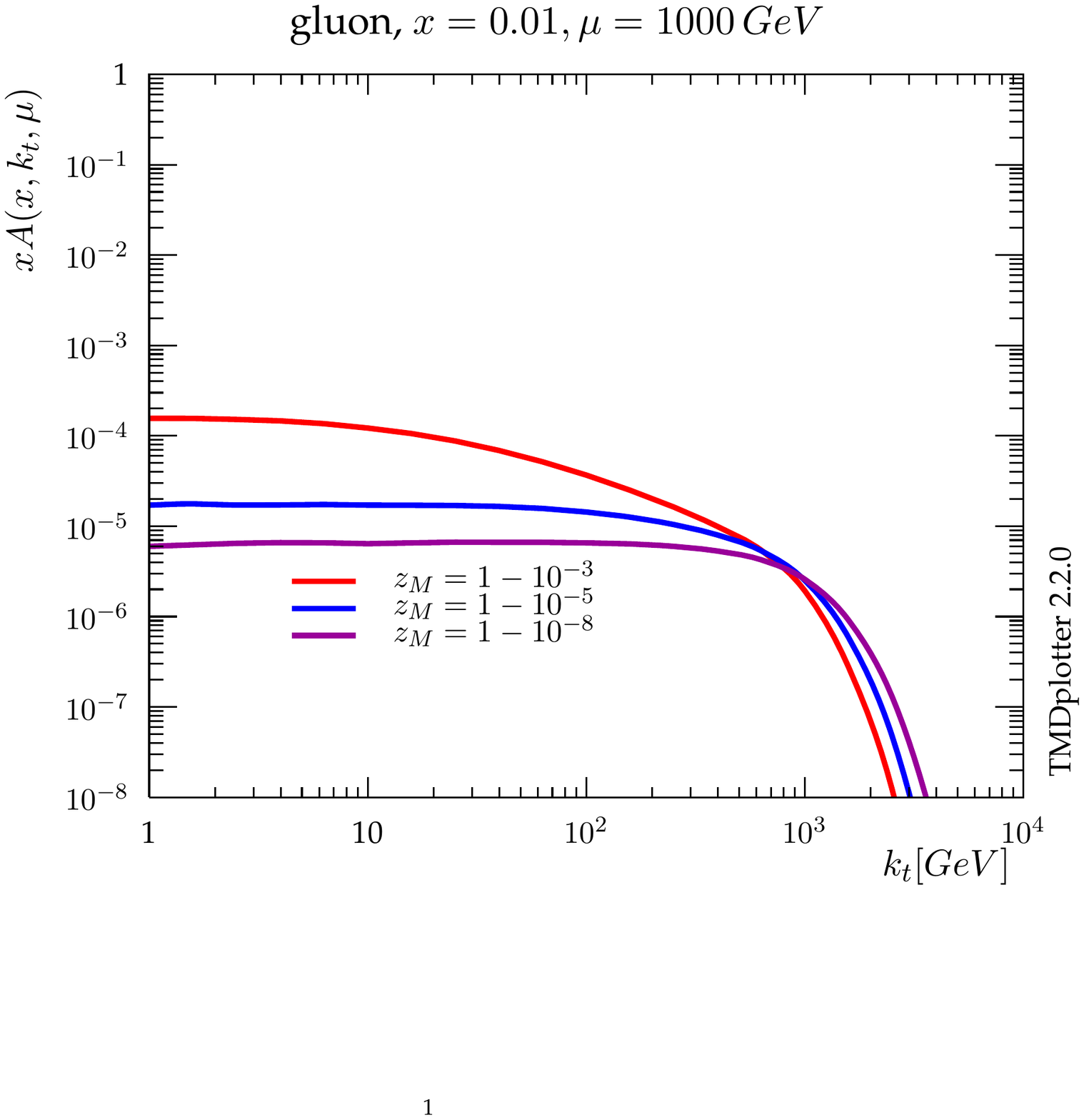}
\caption{\it  Transverse momentum 
gluon  distribution versus $\kt$ 
at $x= 10^{-2} $ and  $\mu=100 
\ {\rm{GeV}}$   
(upper row), $\mu=1000  
\ {\rm{GeV}}$   
(lower row)
for different values of the resolution 
scale 
parameter  $z_{M}=1-10^{-3}, 1-10^{-5}, 1-10^{-8}$:  (left) 
angular ordering;  (right)  transverse momentum ordering.}
\label{fig:fig7}
\end{center}
\end{figure}

 It has been  shown  
 in~\cite{Hautmann:2017xtx} 
that the TMD  distributions, 
unlike the  collinear distributions, 
are strongly influenced by 
the ordering variable in the 
branching. In particular, the cases 
of the transverse momentum ordering (\ref{qtord}) and angular ordering (\ref{angord}) have been examined 
in~\cite{Hautmann:2017xtx} by an explicit calculation, working 
at LO in the strong coupling $\as$. 
We here confirm and extend 
this analysis, working at NLO.   
We illustrate  that 
the same behavior found at LO 
applies at  NLO as well. 

In Figs.~\ref{fig:fig6}  and \ref{fig:fig7}  we apply the 
NLO numerical solution of 
Sec.~\ref{sec:3} and the method 
of Sec.~\ref{subsec:2g}
to study the longitudinal and 
transverse momentum 
dependence of the gluon distribution, 
and its behavior with the soft-gluon 
resolution parameter $z_M$. 
 Fig.~\ref{fig:fig6}  shows 
 the  TMD gluon 
distribution versus the longitudinal momentum 
fraction $x$ 
for different values of 
the resolution parameter, 
$1 - z_{M}=10^{-3}, 10^{-5}, 10^{-8}$.
The curves are plotted for 
 a fixed value 
of transverse momentum  $ k_t \equiv 
| {\bf k} |  = 10 $ GeV, and two values 
of evolution scale, $ \mu = 100 
$~GeV (top panels) and 
$ \mu = 1000 
$~GeV (bottom panels).\footnote{The plots in 
Figs.~\ref{fig:fig6}    and \ref{fig:fig7} 
are produced using the plotting tool 
TMDplotter~\cite{tmdplott,tmdplott16}.} 
On the right  are the results for transverse-momentum ordering; on the left are the results for angular ordering. The 
transverse-momentum ordering, due to the 
effect of 
the non-resolvable region,  
does not lead to results 
independent of $z_M$. 
On the other hand, the angular 
ordering correctly 
takes into account the cancellation 
of non-resolvable emissions 
due  to soft-gluon 
coherence~\cite{Catani:1990rr}, and 
no dependence 
is left on the resolution parameter $z_M$. 

Fig.~\ref{fig:fig7}  shows 
 the  TMD gluon 
distribution versus the transverse momentum 
  $ k_t  $,  at fixed $x =10^{-2}$, 
for different values of 
the resolution parameter, 
$1 - z_{M}=10^{-3}, 10^{-5}, 10^{-8}$. 
As in Fig.~\ref{fig:fig6},  
on the right  are the results for transverse-momentum ordering, and  on the left are the results for angular ordering. 
The plots in Fig.~\ref{fig:fig7}
    illustrate 
as a function of  $ k_t$  the same effect of 
the ordering and behavior in the 
resolution parameter $z_M$ which we have 
seen in the previous figure. 

Analogous behavior to that in 
Figs.~\ref{fig:fig6}  and \ref{fig:fig7} was observed 
at LO in~\cite{Hautmann:2017xtx}. 
   Figs.~\ref{fig:fig6}    and \ref{fig:fig7}     
   show 
  that the 
$z_M$ dependence in 
the transverse momentum 
ordering case cannot be avoided or 
reduced by inclusion of 
NLO 
evolution. It means that 
the  different  orderings  
in Eqs.~(\ref{qtord}),(\ref{angord}) 
should  not be thought of as   different 
factorization schemes, and the results 
in the two cases will  not be related by a change in the factorization scheme.  

It is worth 
 noting that the transverse momentum ordering 
(\ref{qtord}) is widely used 
in a variety of contexts, 
e.g.  in  
low-$x$ physics studies, as it 
results from approximating the exact 
parton-splitting kinematics 
in the region of strongly ordered momentum 
fractions $ z \ll 1$.  The results in 
Figs.~\ref{fig:fig6}  and \ref{fig:fig7} 
    imply  
that 
this approximation will  not be 
valid for observables 
sensitive to the detailed 
structure in transverse momentum 
of the initial state. In particular, they emphasize the 
role of soft-gluon coherence effects 
leading to angular ordering in 
constructing well-defined TMD distributions even at 
low $x$~\cite{mw92,hj-ang,deOliveira:2014cua,Martin:2009ii}.

\begin{figure}[htb]
\begin{center} 
\includegraphics[width=0.45\textwidth, trim=90 180 5 100, clip ]{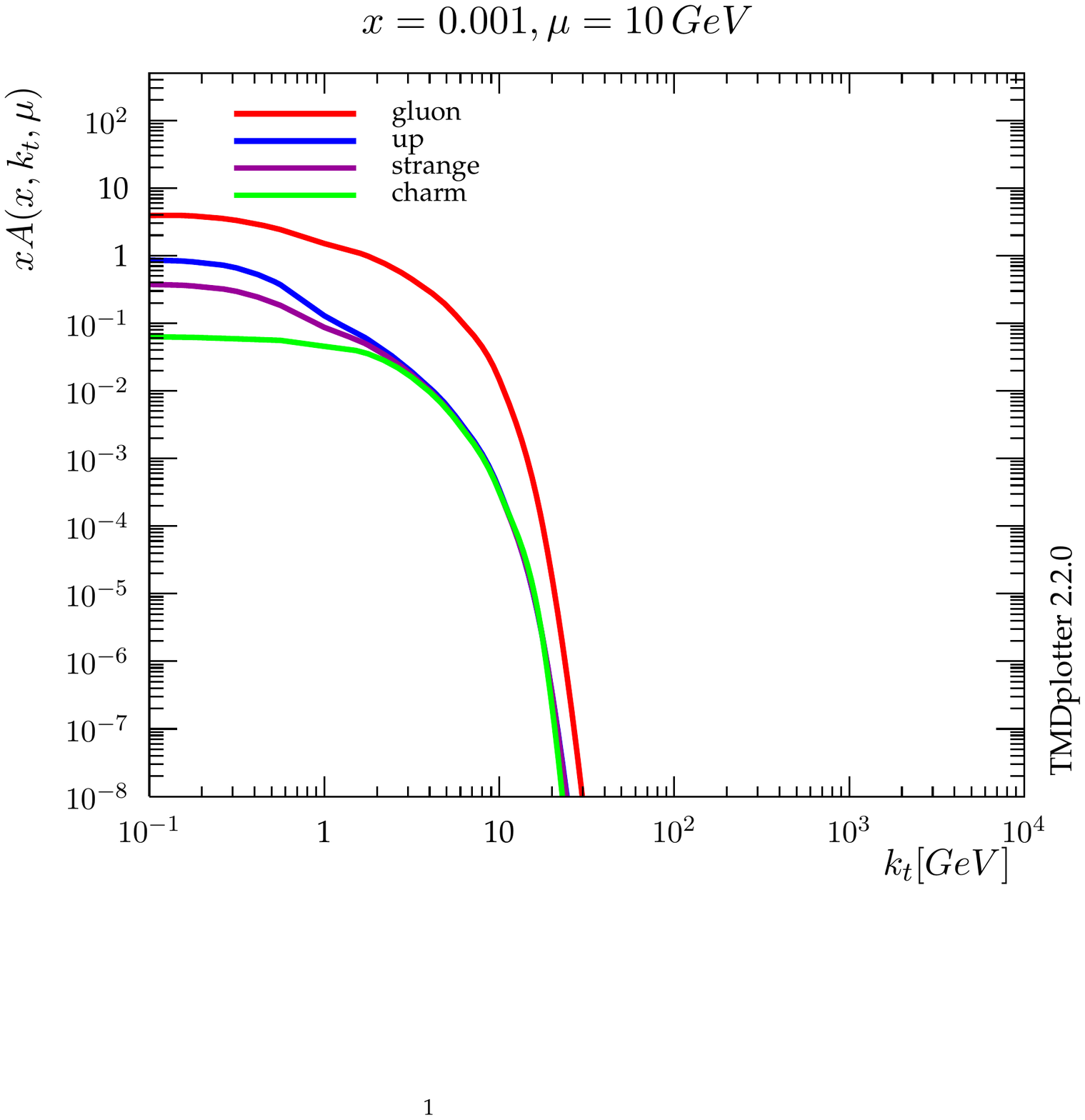}
\includegraphics[width=0.45\textwidth, trim=90 180 5 100, clip ]{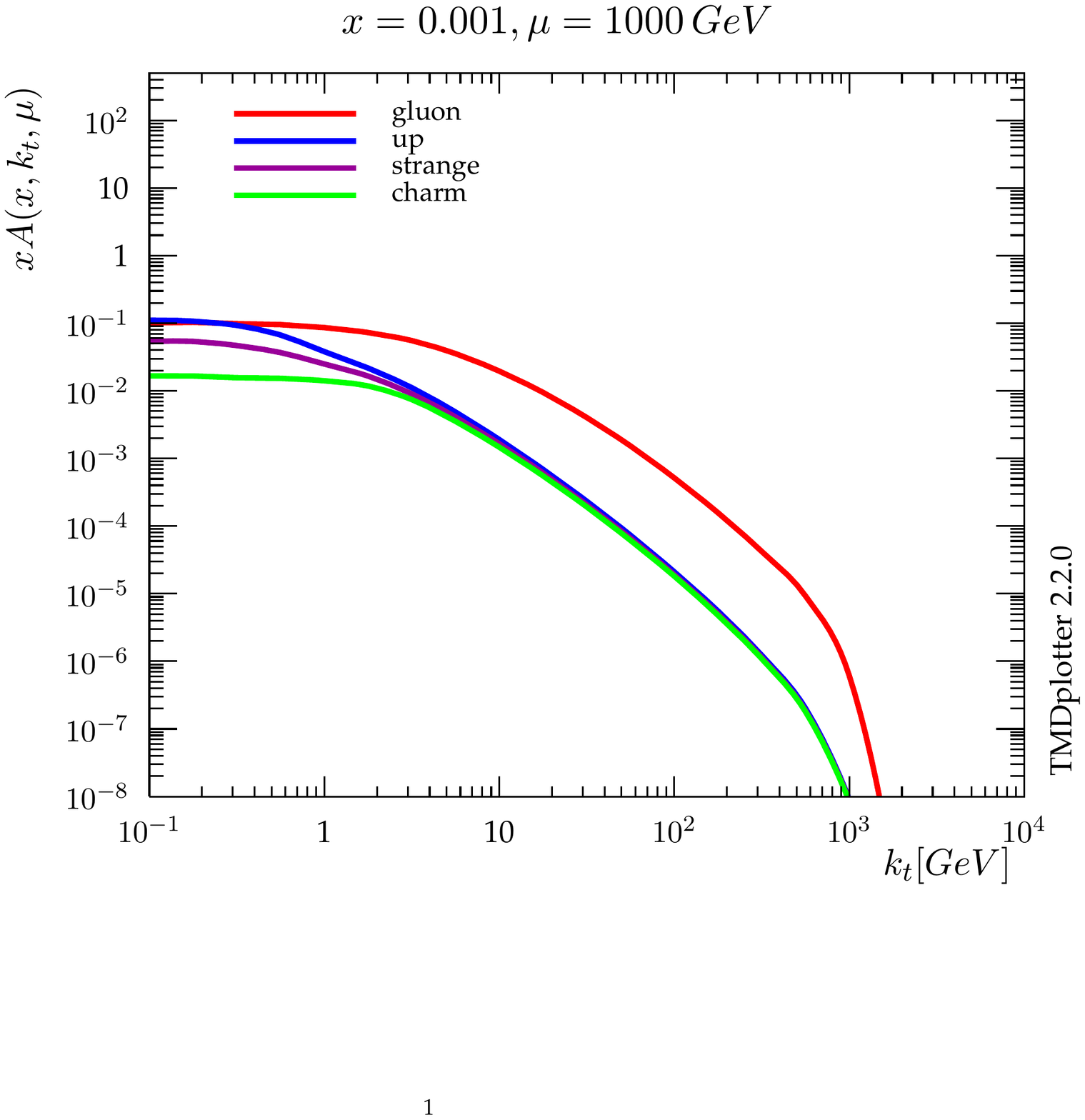}
  \caption{\it Transverse momentum 
distributions  at $x = 0.001$ and  
 evolution scales $\mu = 10 
\ {\rm{GeV}}$ (left), 
$\mu = 1000 \ {\rm{GeV}}$  (right) 
 for different flavors.}
\label{fig:fig8}
\end{center}
\end{figure}

In Fig.~\ref{fig:fig8} we  
illustrate  the flavor decomposition,    
at TMD level,  resulting from perturbative evolution. 
We  plot     
the  TMD distributions obtained from 
the parton branching method for 
different flavors, by applying the evolution 
with appropriate angular-ordering condition.  

To summarize, in this section we have shown that a consistent set of TMD parton distributions, valid over a large range in $x$, $k_\perp$ and  $\mu$ 
can be determined from a parton branching solution of QCD  evolution equations, as long as the soft gluon region is treated appropriately, e.g.~by applying angular ordering  
conditions. We have shown that 
under these conditions 
  the dependence on the  
resolution scale  parameter $z_M$  drops out also for the 
$k_\perp$-distribution,  
provided 
 $z_M$ is large enough, resulting in stable predictions for the evolution 
of TMD parton densities.  

\section{Conclusions} 
\label{sec:6}

Motivated by both conceptual and technical questions on the treatment 
of initial-state kinematics and  distribution  functions in 
QCD parton-shower calculations, 
we have investigated 
parton-branching solutions to QCD  evolution equations. 
We have presented  results  
of   constructing  collinear 
and TMD parton densities from this approach at NLO.

By 
separating 
 resolvable and 
non-resolvable branchings, 
and   
analyzing   
 the role of  the 
soft-gluon resolution scale in the 
evolution, we have 
  proposed a method to   take   
into account simultaneously 
soft-gluon emission in the region 
$z \to 1$  and transverse momentum 
$ {\bf q}_\perp$ recoils in the parton branchings along the QCD cascade. 

This approach is potentially relevant 
for calculations both in collinear 
factorization and in transverse momentum dependent factorization. 
The starting point of the 
approach are the DGLAP equations, 
which are evolution equations  
valid for fully inclusive distributions.  
The method developed in this paper 
provides the branching equations 
which apply at exclusive level. Unlike DGLAP equations, these are 
necessarily sensitive to soft-gluon emission in the infrared region.  We have presented the evolution equations as a function of the soft-gluon resolution scale and the ordering condition. 

The branching equations for TMD 
densities obtained in this paper 
can be compared with existing 
TMD evolution equations: 
the CSS  
equations~\cite{jcc-book,johndave80s}, which apply in 
the low transverse momentum region 
$ q_\perp \ll Q$ (where $Q$ is the high mass scale of the hard scattering)  and can be used in the case of low-$q_\perp$  TMD 
factorization~\cite{css80s};  
and 
the CCFM 
equations~\cite{Catani:1989sg}, which 
apply in the high-energy region 
$\sqrt{s} \gg Q$, and can be used in the case of high-energy TMD 
factorization~\cite{hef}.
We 
have pointed to a 
few of the main differences and similarities in the physics described by these different approaches. 

The 
CSS and CCFM 
approaches 
are designed to achieve high logarithmic accuracy in the resummation of higher-order logarithmic contributions in the 
restricted phase space regions which 
identify their 
domains 
of validity, respectively $ q_\perp 
\to 0$ and $\sqrt{s} \to \infty$. 
In such  approaches 
matching methods are 
required 
to go beyond these restricted 
domains 
and arrive at predictions valid more 
generally (e.g., the 
$Y$-term matching for high $q_\perp$ in CSS, 
and the large-$x$ terms in 
CCFM splitting functions). 
On the other hand, 
the spirit of the approach 
proposed in this paper is to 
provide 
TMD distributions 
which can be applied over a  broad  kinematic 
range 
from  
low to high energies, and from low to high $q_\perp$.   
We 
incorporate consistently 
renormalization 
group evolution, soft-gluon coherence 
and parton branching kinematics.     
The approach is general and,  
although 
in this paper we 
focus on 
longitudinal splitting functions, 
we believe it 
can accommodate 
further 
 dynamical effects such as 
transverse splitting 
functions~\cite{Hautmann:2012sh,Hautmann:2007uw}. 

The formalism of this work implies 
that the soft-gluon resolution scale 
$z_M$ 
depends on the evolution variable 
$\mu$. In the numerical examples 
of this paper 
we have limited ourselves to 
considering fixed values of $z_M$. 
One of the main directions of development of this approach will concern the $\mu$ dependence of 
$z_M$. 

Furthermore, we observe  that, 
while power-suppressed contributions 
of order $ {\cal O } ( 1 - z_M)^n \sim
 {\cal O } (\Lambda_{\rm{QCD}} / \mu)^n $ are beyond the scope of 
the treatment given in this paper, 
logarithmically enhanced contributions 
in $\ln^n   (1 - z_M) $ could 
be taken into account, and 
could be 
related~\cite{Catani:1990rr}
with threshold logarithms 
in production cross sections coupled 
to the parton distributions. 
We regard this as a further potential advantage of 
the  formalism of this work.  

Given the results presented in this 
paper for the  parton 
evolution including the full flavor structure,  
we 
expect this approach 
 to have a wide range of 
applications 
both at low energies and at 
high energies. 

\vspace*{1cm} 

\noindent {\bf Acknowledgments}.    
We thank Z.~Nagy and S.~Jadach for useful discussions. 
FH  acknowledges the support and hospitality of DESY, the Terascale Physics Helmholtz Alliance and the 
DFG Collaborative Research 
Centre SFB 676 
``Particles, Strings and the Early Universe" while part of this work was being done.

\vspace*{1cm} 

\noindent {\Large \bf Appendix A}
\vskip 0.3 true cm 

\noindent 
In this appendix we  report  the 
two-loop coefficients of  the perturbative expansion  
   (\ref{Rab})    for the functions 
$ R_{ab} $ introduced 
in Eq.~(\ref{decompPab}). 
The coefficients can be read 
from the two-loop results 
in~\cite{cfpref1,cfpref2}.

We introduce  the functions 
\begin{equation}
p_{qq}(z)=\frac{2}{1-z}-1-z,
\end{equation}
\begin{equation}
p_{qg}(z)=z^{2}+(1-z)^{2},
\end{equation}
\begin{equation}
p_{gq}(z)=\frac{1+(1-z)^{2}}{z},
\end{equation}
\begin{equation}
p_{gg}(z)=\frac{1}{1-z}+\frac{1}{z}-2+z(1-z),
\end{equation}
and
\begin{equation}
S_{2}(z)=-2\textrm{Li}_{2}(-z)+\frac{1}{2}\ln ^{2}z -2\ln z\ln (1+z)-\frac{\pi^{2}}{6} \; , 
\end{equation}
where the dilogarithm function   is  defined  by 
\begin{equation}
\textrm{Li}_{2}(  y )  =    -  \int_0^y    {{ dt }  \over  t} \ln (1 -t) 
 \; . 
\end{equation}

The two-loop  
contributions 
$R_{ab}^{(1)}$ in   Eq.~(\ref{Rab})  are given 
for quark-gluon and gluon-gluon cases 
by 

\begin{eqnarray}
\label{rgq1} 
R_{gq}^{(1)}(z)&=& C_{F}^{2}\Big[-\frac{5}{2}-\frac{7}{2}z+\Big(2+\frac{7}{2}z\Big)\ln z+\Big(\frac{1}{2}z-1\Big)\ln ^{2}z-2z\ln(1-z)
\nonumber\\
&-& \Big(3\ln(1-z)+\ln^{2}(1-z)\Big)p_{gq}(z) \Big]
+ C_{F}C_{A}\Big[
\frac{28}{9}+\frac{65}{18}z+\frac{44}{9}z^{2}
\nonumber\\ 
&+& 
\Big(-12-5z-\frac{8}{3}z^{2} \Big)\ln z
+(4+z)\ln^{2}z+2z\ln(1-z)+p_{gq}(z) 
\nonumber\\ 
&\times& \Big( -2\ln z\ln(1-z)+\frac{1}{2}\ln^{2}z+\frac{11}{3}\ln(1-z)+\ln^{2}(1-z)-\frac{\pi^{2}}{6}+\frac{1}{2} \Big)
\nonumber\\
&+&  p_{gq}(-z)S_{2}(z)\Big]
 + C_{F}T_{R}N_{f}\Big[-\frac{4}{3}z-\Big( \frac{20}{9}+\frac{4}{3}\ln(1-z)\Big)p_{gq}(z) \Big],
\end{eqnarray}

\begin{eqnarray}
\label{rqg1} 
 R_{qg}^{(1)}(z)&=&{1\over 2} C_{F} T_{R}  \Big[
4-9z+(-1+4z)\ln z+(-1+2z)\ln^{2}z+4\ln(1-z)
\nonumber\\
&+&\Big(-4\ln z \ln(1-z)+4\ln z +2\ln^{2}z-4\ln(1-z)
+2\ln^{2}(1-z)    
\nonumber\\
&-& \frac{2}{3}\pi^{2}+10\Big)  p_{qg}(z)
\Big]  + {1\over 2} C_{A} T_{R}  \Big[
\frac{182}{9}+\frac{14}{9}z+\frac{40}{9z}+\Big(\frac{136}{3}z-\frac{38}{3} \Big)\ln z  
\nonumber\\ 
&-& 4\ln(1-z) -   ( 2+8z)\ln^{2}z  +  \Big( -\ln ^{2}z +\frac{44}{3}\ln z -2\ln^{2}(1-z)  
\nonumber\\
&+& 4\ln(1-z)+\frac{\pi^{2}}{3}-\frac{218}{9}\Big)   p_{qg}(z)  + 2p_{qg}(-z)S_{2}(z) \Big]
\end{eqnarray}

and 

\begin{eqnarray}
\label{rgg1}
R_{gg}^{(1)}(z)&=&C_{F}T_{R}N_{f} \Big[
-16+8z+\frac{20}{3}z^{2}+\frac{4}{3z}+(-6-10z)\ln z +(-2-2z)\ln ^{2}z
\Big]
\nonumber\\
&+&C_{A}T_{R}N_{f}\Big[ 2-2z+\frac{26}{9}z^{2}-\frac{26}{9z}-\frac{4}{3}(1+z)\ln z-\frac{20}{9}\Big( \frac{1}{z}-2+z-z^{2}\Big) \Big]
\nonumber\\
&+&C_{A}^{2}\Big[ \frac{27}{2}(1-z)+\frac{67}{9}\Big(z^{2}-\frac{1}{z}\Big)+\Big(-\frac{25}{3}+\frac{11}{3}z-\frac{44}{3}z^{2} \Big)\ln z 
\nonumber\\
&+&4(1+z)\ln^{2}z+2p_{gg}(-z)S_{2}(z)+\left(-4\ln z \ln(1-z)+\ln^{2}z \right)p_{gg}(z)
\nonumber\\
&+& \Big(\frac{67}{9}-\frac{\pi^{2}}{3}\Big)\Big(\frac{1}{z}-2+z-z^{2}\Big)
\Big]  . 
\end{eqnarray}

The two-loop  
contributions 
$R_{ab}^{(1)}$ for the 
 non-singlet case, 
in the notation of Eq.~(\ref{adpr}),   are given by 

\begin{equation}
\label{rqqbarns1}
R_{q\overline{q}}^{NS(1)}(z)=C_{F}\Big(C_{F}-\frac{C_A}{2} \Big)\Big[2p_{qq}(-z)S_{2}(z)+2(1+z)\ln z+4(1-z) \Big] , 
\end{equation}

\begin{eqnarray}
\label{rqqns1}
R_{qq}^{NS(1)}(z)&=&C_{F}^{2}\Big[
-\Big( 2\ln z \ln (1-z)+\frac{3}{2}\ln z\Big)p_{qq}(z)-\Big(\frac{3}{2}+\frac{7}{2}z\Big)\ln z - \frac{1}{2}(1+z)\ln^{2}z   
\nonumber\\ 
&-& 
   5(1-z)
\Big]
+C_{F}C_{A}\Big[ \Big(\frac{1}{2}\ln ^{2}z+\frac{11}{6}\ln z\Big)p_{qq}(z)
-(1+z)\Big(\frac{67}{18}-\frac{\pi^{2}}{6} \Big)   + (1+z) 
\nonumber\\ 
&\times& \ln z 
 +  \frac{20}{3}(1-z)
 \Big]
 +C_{F}T_{R}N_{f}\Big[-\frac{2}{3}\ln z p_{qq}(z) +\frac{10}{9}\left(1+z \right)-\frac{4}{3}(1-z) \Big]  . 
\end{eqnarray}

By defining   the 
linear combination of  the splitting functions in 
Eq.~(\ref{adpr})
\begin{equation}
\label{adsingl}
P_{qq} 
= P_{qq}^{NS} + P_{q{\bar q}}^{NS} + N_f (P_{qq}^{S} +  
P_{q{\bar q}}^{S})  \;\; , 
\end{equation}
which controls the 
evolution of  the singlet  quark   distribution coupled 
to gluons, 
the corresponding two-loop  
 contribution to the functions 
 $R_{ab}$ in   Eq.~(\ref{Rab})   
  is given by 
 
\begin{eqnarray}
\label{rqqsing1}
R_{qq}^{(1)}(z)&=&C_{F}^{2}\Big[
-1+z+\Big( \frac{1}{2}-\frac{3}{2}z \Big)\ln z-\frac{1}{2}\Big( 1+z\Big)\ln^{2}z+2p_{qq}(-z)S_{2}(z)
\nonumber\\ 
&-& 
\Big(\frac{3}{2}\ln z +2\ln z \ln(1-z)\Big)p_{qq}(z)\Big]
\nonumber\\ 
&+& C_{F}C_{A}\Big[ \frac{14}{3}(1-z)-p_{qq}(-z)S_{2}(z)+\Big(\frac{11}{6}\ln z +\frac{1}{2}\ln^{2} z \Big)p_{qq}(z)
\nonumber\\ 
&-&(z+1) 
\Big( \frac{67}{18}-\frac{\pi ^{2}}{6}\Big) \Big] 
+  C_{F}T_{R}N_{f}
\Big[
-\frac{16}{3}+\frac{40}{3}z+\Big(10z+\frac{16}{3}z^{2}+2 \Big)
\nonumber\\ 
 &\times & \ln z - 
  \frac{112}{9}z^{2}+\frac{40}{9z}-2(1+z)\ln ^{2}z
 - \frac{2}{3}\ln z p_{qq}(z)+\frac{10}{9}(z+1)   \Big]. 
\end{eqnarray}

\end{document}